\documentclass[12pt]{article}
\usepackage{amsmath,amssymb}
\begin{document}

\thispagestyle{empty}
\begin{flushright}
ROM2F/2003/16
\end{flushright}
\vspace{1.2cm}
\begin{center}
{\Large {\bf On two-fermion BMN operators}}
\\
\vspace{0.6cm} {Burkhard Eden} \\ \vspace{0.6cm} {{\it
Dipartimento di Fisica, \ Universit{\`a} di Roma \ ``Tor
Vergata''}} \\  {{\it I.N.F.N.\ -\ Sezione di Roma \ ``Tor
Vergata''}} \\ {{\it Via della Ricerca  Scientifica, 1}}
\\ {{\it 00133 \ Roma, \ ITALY}} \\
{{\it e-mail: burkhard.eden@roma2.infn.it}} \\
\end{center}
\vspace{0.6cm}

\begin{abstract}

We show how to determine the lowest order mixing of all scalar with two-fermion
two impurity BMN operators in the antisymmetric representation of $SO(4)$.
Differentiation on harmonic superspace allows one to derive two-loop anomalous
dimensions of gauge invariant operators from this knowledge: the value for
the second anomalous correction to the dimension is essentially the square
of the two-fermion admixture. The method effectively increases the loop order
by one. For low $J$ we find agreement to all orders in $N$ with results
obtained upon diagonalisation of the ${\cal N}=4$ dilation operator.

We give a formula for the generalised Konishi anomaly and display
its role in the mixing. For $J=2$ we resolve the mixing up to
order $g^2$ in the singlet representation. The sum of the anomaly
and the naive variation of the leading two-fermion admixtures to
the singlets is exactly equal to the two-fermion terms in the
antisymmetric descendants.

\end{abstract}

\newpage

\setcounter{page}{1}

\section{Introduction}
\label{intro}

The AdS/CFT correspondence \cite{eins} in its strong form has the
drawback that the string side of the duality is virtually
inaccessible to calculation. More recently, a special limit of the
underlying geometry has been considered \cite{ppwaves}, in which
the string theory becomes solvable \cite{mets}. The article
\cite{bmn} established a field theory dual. String states are
related to composites of very many copies of a given scalar field
of ${\cal N}=4$ SYM with a few other elementary fields, commonly
termed ``impurities".

This work addresses once again the set of operators with two
impurities: we focus on the mixing between operators made out of
only scalar fields and scalar operators with two-fermion
impurities. In the BMN proposal the $SU(4)$ $R$-symmetry of the
${\cal N} = 4$ theory is broken to $U(1)_J \otimes SO(4)$. The
chosen scalar field $Z$, say the field $\phi_1$ in the ${\cal N} =
1$ formulation of the theory, is charged under $U(1)_J$ but does
not transform under the $SO(4)$ factor. The impurities are neutral
but rotate under $SO(4)$. Two impurities may carry a singlet, an
antisymmetric and a symmetric traceless representation of $SO(4)$.

We do not enter into the subtleties of the actual limit,
although the study is certainly motivated by the BMN proposal. In
the full ${\cal N}=4$ theory with unbroken $SU(4)$ R symmetry the
two impurity BMN operators are highest weights of an $[0,J,0]$, a
$[2,J,0]$ and a $[2,J,2]$ irrep of $SU(4)$, respectively. The
operators are made out of many elementary fields, which can be
arranged into traces of the associated gauge group generators in
many different ways. Since all these objects have the same naive
conformal dimension one has to disentangle the operator mixing,
i.e. to find operators with well-defined conformal dimension in
the quantum theory.

Operator mixing in ${\cal N}=4$ is difficult to solve exactly even
at the lowest order in the coupling constant \cite{op1,op2}. The
large $N$ limit provides a natural simplification. The original
work \cite{bmn} gives a solution for this case. Later on much
effort has been dedicated to determining subleading (in $N$)
corrections to the one-loop anomalous dimensions
\cite{danetal,plefkaetal} and there is even a two-loop calculation
\cite{grossetal}. Degeneracy of the anomalous dimensions of various
types of operators has led to the conjecture that they belong to
supermultiplets, see \cite{gursoy} for vector operators and a comment on
two-fermion operators, and \cite{beisert} for the all-scalar composites.
According to the latter article the highest weight states
are the singlets, the antisymmetric and symmetric operators are
descendants.

For $J=2$, this had first been observed in \cite{op1}. A part of this article
is dedicated to demonstrating how the structure of the mixing problem
in the $J=2$ example generalises to the whole class of operators.

From an ${\cal N} = 4$ perspective we are dealing with scalar
multiplets of $SU(2,2|4)$ which carry an $SU(4)$ representation
$[0,J,0]$ and have naive scaling weight $\Delta = J+2$. Such
operators are semishort \cite{bps3} in free field theory and may
become long in the interacting case.

Interestingly, the two-impurity all-scalar BMN operators are the highest
weights of four multiplets that are separate in free field theory
($[0,J,0],\Delta=J+2$; \, $[2,J,0], \Delta=J+3$ and the conjugate
representation, $[2,J,2],\Delta=J+4$), but that merge if
interactions are switched on \cite{howe}: the descendant
structure \cite{beisert} indeed derives from the commutator term in the
supersymmetry transformations, which comes with the Yang-Mills
coupling constant.

But there is a second effect, which we illustrate by an example:
the two operators ${\cal K}_1 = (\phi_I \bar \phi^I)$ and the
lowest component of the stress-energy tensor ${\cal Q}_{20}$ are
orthogonal, because they are in different representations. Under
classical supersymmetry they have descendants in the same
representation, namely
\begin{equation}
{\cal K}_{10} \, = \, g (Z[\phi_2,\phi_3]) \, \qquad {\cal Q}_{10}
\, = \, (\lambda^\alpha \lambda_\alpha) + 4 g (Z[\phi_2,\phi_3])
\end{equation}
(up to scaling) which are clearly not orthogonal to order $g^2$.
The explanation is that we have omitted an ``anomalous'' part of the
supersymmetry variation of ${\cal K}_1$: the correct descendant is
\begin{equation}
{\cal K}_{10} \, = \, g (Z[\phi_2,\phi_3]) + \frac{g^2 N}{32
\pi^2} (\lambda^\alpha \lambda_\alpha) \, .
\end{equation}
It was shown in \cite{konishi} how the missing piece of the descendant
can be derived in a graph calculation. In point splitting
regularisation one must insert the gauge connection between two
elementary fields that are not at the same point. It is the
supersymmetry variation of the connection that accounts for the
two-fermion part of the descendant. The two-fermion piece became
known as the ``Konishi anomaly''.

In a separate paper we will present a graph calculation concerning
the analogous anomaly for $[0,J,0]$ operators with weight
$\Delta=J+2$. In BMN inspired notation the highest weight of such
an operator can be a combination of
\begin{equation}
{\cal O}_I \, = \, Z^J \phi_a \bar \phi^a \, , \qquad {\cal
O}_{III} \, = \, Z^{J+1} \bar Z
\end{equation}
($a =2,3$) in some arrangement of the fields into gauge group
traces. We preempt the result of the exercise: the anomaly of the BMN
singlets is correctly reproduced by the functional differential
operator\footnote{M. Bianchi independently derived a more general
formula \cite{bianchi}.}
\begin{equation}
{\cal F}_K \, = \, - \frac{1}{16} \, \frac{g^2}{4 \pi^2} \,
\Biggl( \Bigl( \lambda^\alpha \Bigl[ \lambda_\alpha,
\frac{\delta}{\delta \phi_I} \Bigr] \frac{\delta}{\delta
\bar \phi^I} \Bigr) + (\phi \leftrightarrow \bar \phi) \Biggr)
\label{Massimo}
\end{equation}
where $I=1,2,3$ (so the operator does act on $Z,\bar Z$).
Supersymmetry and orthogonality considerations much like in the example
above suffice to fix the lowest order two-fermion admixtures to the all
scalar BMN's. Order $g^2$ orthogonality then yields the $g^2$ mixing, too.

Next, superspace two-point functions of primary operators and
descendants have different normalisations, because the descendant
is usually obtained by a differentiation which brings out factors
depending on the dimension of the operator \cite{anselmi}. We consider
the standard gauge invariant BMN operators as opposed to
\cite{forzaitalia} which is concerned with gauge non-invariant
composites. It turns out that the differentiation trick when applied to the
BMN operators relates the two-loop anomalous dimensions rather directly
with the two-fermion admixtures of the antisymmetric descendants.
For $J=0,1,2$ and gauge group $SU(N)$ we work out the two-loop anomalous
dimensions from a one-loop calculation. We agree to all orders in $N$
with results obtained from the two-loop dilation operator of \cite{dilat}.
On the other hand the two-loop dilation operator approach reproduces the
results of the $g^4$ graph calculation \cite{grossetal}.

The idea of the dilation operator grew out of the recent work about
spin chains realised by ${\cal N} = 4$ operators \cite{spin}, and references
therein. Integrability of the spin chain enables the authors of \cite{dilat} to
predict anomalous dimensions up to three and four loops. The main aim of
this article is to advocate the superspace differentiation method
as a possible way to check these claims, since it effectively increases
the loop order by one.

Finally, we demonstrate the validity of (\ref{Massimo}) by
matching the double-fermion admixture in the antisymmetric
representation with the sum of the anomaly and the classical
variation of a two-fermion addition in the singlet. In particular,
the protected weight four double trace operator ${\cal D}_{20}$
\cite{d20,esok,kon} has a non-zero anomaly. The multiplet can be short
only because there is a cancellation against another term.

The formula for the generalised Konishi anomaly is intereresting in its
own right. It would be fascinating to make contact with \cite{dougwit} which
derives an anomaly for a similar set of operators in an ${\cal N} = 1$ setting.

\subsection{Plan of the paper}
\label{plan}

In Section (\ref{bmnops}) we address the diagonalisation of BMN
operators with two scalar impurities in each of the three possible
$SO(4)$ representations (singlet, antisymmetric and symmetric
traceless). Although we do not aspire to resolve the mixing for
arbitrary $N$, we give bases for one-loop protected and
unprotected operators in each of the three representations and
demonstrate the descendant structure for general $N$ and $J$. We
show that the one-loop mixing matrix\footnote{To be more precise,
we should talk about the tree-level and single logarithm parts of
``the matrix of two-point functions''. Throughout the paper we
avoid this more correct but rather clumsy nomenclature in favour
of ``mixing matrix''.} of the singlet operators equals the
tree-level mixing of their antisymmetric descendants. The
situation persists between the antisymmetric operators and their
symmetric descendants.

Section (\ref{suptwopt}) explains the aforementioned equivalence
of mixing matrices on the basis of differentiation of abstract
two-point functions on ${\cal N} = 4$ harmonic superspace. The
absence of descendants for the one-loop protected operators is
shown to take the form of a shortening condition of the
``semishort" type \cite{bps3}.

In Section (\ref{fermimp}) we discuss mixing with operators
involving fermion impurities in the antisymmetric representation.

In Section (\ref{opmix}) we fix the operator mixing up to order
$g^2$ in the antisymmetric and symmetric representations,
for $J=0,1,2$ and $SU(N)$ gauge group. We derive the
two-loop anomalous dimensions by differentiation on superspace.

In Section (\ref{dilat}), we check consistency of our results with
the diagonalisation of the two-loop dilation operator.

Finally, in Section (\ref{singsec}) we fix the operator mixing
through $g^2$ in the singlet. We check our formula for the anomaly
and re-derive our equation for the two-loop anomalous dimensions.

In two appendices we discuss the ${\cal N} = 4$ supersymmetry
transformations and the $SU(4)\rightarrow SO(4) \times U(1)$
branching relevant in the BMN limit, and give technical details of
the calculations of Section (\ref{bmnops}).

\section{BMN operators with two impurities}
\label{bmnops}
Throughout the paper it is assumed that for each value of $J$ the rank
of the gauge group $N$ is high enough for all operators to be independent.

We distinguish two classes of operators: type I has both
impurities in the same gauge trace, type II has the impurities in
different gauge traces. We shall study the tree-level and one-loop
mixing of the charge $J$ objects
\begin{eqnarray}
{\cal O}^{(J_0,J_1|J_2...J_k)}_{I, ab} & = &
(\phi_a Z^{J_0} \phi_b Z^{J_1-J_0}) \prod_{i=2}^n (Z^{J_i})
\\
{\cal O}^{(J_0,J_1|J_2...J_k)}_{II, ab} & = & (\phi_a
Z^{J_1-J_{0}}) (\phi_b Z^{J_0}) \prod_{i=2}^n (Z^{J_i})
\end{eqnarray}
with total $U(1)_J$ charge $J=\sum_{i=1}^{n} J_{i}$ and $J_0\le
J_1$ ($J_0 \neq 0$ for type II in $SU(N)$). To save space we have
denote traces with parantheses $()$. The impurities $\phi_a,\phi_b$ can
be any of $\phi_2, \phi_3, \bar \phi^2, \bar \phi^3$. Operators
involving fermion or gauge field impurities decouple from these at
order $g^2$, where g is the YM coupling.

The one-loop combinatorics has the surprising feature of not touching upon
the factor $\prod_i Z^{J_i}$. We will therefore often avoid writing out
the product and rather use the abbreviation $\Pi_Z$ in most
formulae.

Since resolving the mixing has proven to be much more intricate
than originally expected we will mainly focus on identifying
protected operators and/or decoupling of classes of operators from
other classes. To distinguish the representations, we use $st$ to
denote symmetric tensors, $as$ to denote antisymmetric tensors and
we put $sin$ for singlets.

We use the ${\cal N} = 1$ formalism in Euclidean signature. The
tree-level two-point functions of charge $J$ singlet operators
have coordinate dependence
\begin{equation}
\langle \, {\cal O}_{sin} \, {\cal O}_{sin}^\dagger \,
\rangle_{g^0} \, = \, \frac{1}{(4 \pi^2 \, x^2_{12})^{J+2}} \; .
\label{treex}
\end{equation}
What is more, there is only one one-loop superspace
integral.\footnote{There is a caveat here: the ${\cal N}=1$
formalism needs a connection between chiral and antichiral fields.
For the singlets there is a class of graphs involving a YM line
emanating from the connection. These always seem to combine with
another class of YM exchange to make the above statement true. We
rely on this assumption.} Its $\theta, \, \bar \theta = 0$
component yields
\begin{equation}
\langle \, O_{sin} \, O_{sin}^\dagger \, \rangle_{g^2} \, = \,
\frac{1}{(4 \pi^2 \, x^2_{12})^J} \, \frac{g^2}{(4 \pi^2)^4} \int
\frac{d^4x_5}{x_{15}^2 \, x_{25}^2} \, = \, \frac{1}{(4 \pi^2 \,
x^2_{12})^{J+2}} \, \frac{g^2}{4 \pi^2} \, \frac{1}{2} \Bigl(
\ln(x^2_{12}) + \alpha \Bigr) \, . \label{loopx}
\end{equation}
We have not indicated the divergence in the integral. It has to be
cancelled by renormalisation of the correctly diagonalised
operators which introduces the scheme dependent constant $\alpha$
behind the logarithm.

The difficulty lies in the combinatorics for $U(N)$ or $SU(N)$
gauge group. In this section we do not explicitly evaluate the
combinatorics, but rather present proofs based on only a few Wick
contractions, employing the rules
\begin{eqnarray}
(T^a A)(T^a B) & = & (A B) - \frac{c_0}{N} (A)(B) \, ,\\ (T^a A
T^a B) & = & (A)(B) - \frac{c_0}{N} (A B) \, ,
\end{eqnarray}
where $( )$ denotes a trace and $c_0 = 0,1$ in $U(N)$ and $SU(N)$,
respectively.

The scalar fields are to be contracted using the ``propagator"
$\langle \phi_I^a \, \bar \phi^{J \, b} \rangle \, = \, \delta^{ab} \,
\delta_I^J$. The idea is to contract the impurities and leave $Z$
and $\bar Z$ untouched \cite{plefkaetal}. Chain sums occuring in the
one-loop calculations are collected into correlators with one more
occurrence of $Z, \bar Z$. In the one-loop calculations all trace
terms with $c_0$ actually cancel; the formulae hold for $U(N)$ as
well as $SU(N)$.

We will move part of the discussion to Appendix (\ref{technic}) in
order to avoid blurring the conclusions in the main text.

\subsection{The antisymmetric representation}

The tree-level mixing between a type I and a type II object
(so far without explicit symmetrisation) is given in
(\ref{app11}). It is invariant under both $J_{0} \leftrightarrow
J_{1}-J_{0}$ and $\bar{J}_{0} \leftrightarrow
\bar{J}_{1}-\bar{J}_{0}$. Consequently, the mixing vanishes if one
of the two operators is antisymmetrised w.r.t. exchange of the two
impurities. Thus the antisymmetric type I operators decouple at
tree-level from type II.

The type I / type I and type II / type II one-loop mixing is given
in formulae (\ref{app12}) and (\ref{app13}), respectively. From
the symmetry under exchange of the impurities we deduce that antisymmetric
type II operators have vanishing one-loop mixing with anything else.
They are one-loop protected.

In the antisymmetric representation there is therefore a very
clear-cut criterion: the potentially unprotected operators are of
type I. They are automatically tree-orthogonal to the (one-loop)
protected operators. Note that any antisymmetric type I operator
is a sum of commutators and can be obtained as a supersymmetry
variation of some singlet.

\subsection{The symmetric traceless representation}

Most conveniently we choose a representative with twice the same
chiral impurity field, say $\phi_2$. The only non-vanishing graphs
involve a matter exchange corresponding to the effective vertex
\cite{plefkaetal}
\begin{equation}
:([\bar{Z},\bar \phi^2],[Z, \phi_2]): \, .
\end{equation}
When looking for protected operators we can restrict our attention
to one ``half'' of the graphs: the contraction of the antichiral
matter vertex on the operator has an open $\bar \phi^3$ leg. The
chiral matter vertex contracted with the conjugate operator at the
other end of the two-point function similarly has an open $\phi_3$
leg, which is its only connection to the first half. The vanishing
of the first half of the two-point function is a sufficient
condition for an operator to be one-loop protected.

We are therefore led to consider linear combinations
\begin{equation}
{\cal L} = \sum_f c_f \, {\cal O}_{st,\; I}^{f} + \sum_h d_h \,
{\cal O}_{st, \; II}^{h}
\end{equation}
of type I and type II symmetric operators, whose contraction on
the antichiral matter vertex vanishes. Explicit calculations for
low $J$ show that these exhaust more than one half of the total
space of operators.

Let us sharpen this statement. It is a quick calculation to check
that all terms in the contraction of the antichiral matter vertex
on any type I or II symmetric operator have the gauge trace
structure of antisymmetric type I operators of charge $J-1$, with
impurities $\phi_2$ and $\bar \phi^3$. Since the contraction
operation is linear, its kernel --- the protected operators arising
in this way --- has as its dimension the number of all symmetric
charge $J$ operators minus the dimension of the image. The
remaining, potentially unprotected operators ${\cal O}_u$ must be
tree-level orthogonal to this set of protected linear
combinations. The dimension of this space is exactly that of the
image of the contraction operation. It is therefore smaller or
equal to the number of antisymmetric charge $J-1$ operators.

The tree-orthogonality condition is:
\begin{equation}
\langle {\cal L} \; {\cal O}_u^\dagger \rangle \, = \, 0
\label{ortho}
\end{equation}
We will now prove that descendants of charge $J-1$ antisymmetric
operators have this property. Take a representative
\begin{equation}
{\cal O}_{as, \; I} = \Pi_Z (\phi_2 Z^{J_{0}} \bar \phi^3
Z^{J_{1}-J_{0}-1}) - (J_{0} \leftrightarrow J_{1}-J_{0}-1)
\end{equation}
and apply the supersymmetry variation\footnote{The commutator term
in the supersymmetry transformations (\ref{dPhi}) comes with a
factor of $g$, which is omitted here for reasons of simplicity.}
\begin{equation}
(\bar \delta^1)^2 \, \bar \phi^I \rightarrow \frac{1}{2}
\epsilon^{IJK} [\phi_J, \phi_K] \; , \qquad (\bar \delta^1)^2 \,
\phi_I = 0 \, .
\end{equation}
The symbol refers to the double application of a certain
supercharge of the ${\cal N} = 4$ theory, see Appendix
(\ref{appnot}). We find
\begin{equation}
(\bar \delta^1)^2 \, {\cal O}_{as, \; I} = 2 \Pi_Z (\phi_2
Z^{J_{0}+1} \phi_2 Z^{J_{1}-J_{0}-1}) - 2 \Pi_Z (\phi_2
Z^{J_{1}-J_{0}} \phi_2 Z^{J_{0}}) \, , \label{descentAnti}
\end{equation}
a difference of symmetric charge $J$ type I operators. We can
write the double supersymmetry transformation as a contraction on
a chiral vertex. Then the tree-orthogonality condition
(\ref{ortho}) becomes
\begin{equation}
\langle {\cal L} \; ([\bar \phi^2, \bar{Z}] \bar \phi^3) {\cal
O}_{as, \; I}^\dagger \rangle = 0
\end{equation}
and is automatically fulfilled since the contraction of the
antichiral vertex on $L$ is zero by definition.

Now, any charge $J-1$ antisymmetric type I operator has such a
descendant, whereas $(\bar \delta^1)^2$ yields zero when acting on
a type II operator. Thus we have shown that a basis for the
symmetric charge $J$ operators carrying one-loop anomalous
dimension is given by the descendants of the antisymmetric charge
$J-1$ type I operators. In particular, these are simply
differences of symmetric type I operators.

On the other hand, the coefficients in the protected linear
combinations are $N$ dependent and we did not obtain them in
closed form.

\subsection{The singlets}

For technical convenience we introduce $SO(4)$ singlets: let us
define
\begin{eqnarray}
{\cal O}^{(J_0,J_1|J_2 \ldots J_k)}_{sin, \, I} & = & (\phi_a
Z^{J_0} \bar \phi^a Z^{J_1-J_0}) \prod_{i=2}^k (Z^{J_i}) + (J_0
\leftrightarrow J_1 - J_0) \, ,
\\ {\cal O}^{(J_0,J_1|J_2 \ldots J_k)}_{sin, \, II} & = & (\phi_a
Z^{J_0})(\bar \phi^a Z^{J_1-J_0}) \prod_{i=2}^k (Z^{J_i}) + (J_0
\leftrightarrow J_1 - J_0) \, ,
\\ {\cal O}^{(J_1|J_2...J_k)}_{sin, \, III} & = & (\bar{Z} Z^{J_1})
\prod_{i=2}^k (Z^{J_i}) \, ,
\end{eqnarray}
with $a =2,3$ in the first two operators. Tree-level
orthogonalisation w.r.t.\ one-loop protected operators organises
the $SO(4)$ singlets into components of various $SU(4)$
representations, see below.

In order to identify protected operators we consider directly the
one-loop mixing matrix of the singlets. Details of the calculation
are given in Appendix (\ref{technic}). We find that type II
operators are protected. The non-vanishing pieces of the mixing
matrix are type I / I, I / III, and III / III and are given in
(\ref{app14}),(\ref{app15}) and (\ref{app16}), repectively. The
matrix is singular and it is not hard to find the zero
eigenvectors. For each type III operator there is a protected
linear combination with a set of type I operators with the same
$\Pi_i (Z^{J_i})$. As a basis for the unprotected operators we may
choose type I. The tree-level orthogonalisation w.r.t. the
protected structures involves coefficients with rather non-trivial
$N$ dependence and we did not obtain a result in closed form.

The expressions for the one-loop mixing of the various operators
are exactly the negative of the tree-level mixing of their
descendants under $(\bar \delta^1)^2$. This observation extends to
the type II operators: they vanish under the supersymmetry
variation and correspondingly the one-loop contribution to their
two-point function with any other operator is zero.

Let us now return to the $SU(4)$ picture. The operators pick up a
$Z \bar{Z}$ piece:
\begin{eqnarray}
{\cal O}_{sin, \, I}^{({J}_0,{J}_1|\ldots)} & = & \Pi_Z \Bigl(
(\phi_2 Z^{J}_0 \bar \phi^2 Z^{{J}_1-{J}_0}) + (\phi_3 Z^{J_0}
\bar \phi^3 Z^{{J}_1-{J}_0}) + (Z^{{J}_1+1} \bar Z) \nonumber
\\ & & \qquad + (J_0 \leftrightarrow J_1-J_0) \Bigr) \, , \\
{\cal O}_{sin, \, II}^{({J}_0,{J}_1|\ldots)} & = & \Pi_Z \Bigl(
(\phi_2 Z^{J_0}) (\bar \phi^2 Z^{{J_1}-{J_0}}) + (\phi_3 Z^{J_0})
(\bar \phi^3 Z^{{J_1}-{J_0}}) + (Z^{{J_0}+1}) (Z^{{J_1}-{J_0}}
\bar{Z}) \nonumber \\ & & \qquad + ({J_0} \leftrightarrow
{J_1}-{J_0}) \Bigr) \, .
\end{eqnarray}
The property of protectedness of the type II operators is now
lost. As a guideline for identifying the one-loop protected
operators we use the criterion that they be annihilated by $(\bar
\delta^1)^2$. We introduce a change of basis
\begin{eqnarray}
\tilde {\cal O}_{sin, \, I}^{({J}_0,{J}_1|\ldots)} & = &
\sum_{f=0}^{J_0} {\cal O}_{sin, \, I}^{(f,{J}_1|\ldots)} -
\frac{J_0+2}{J_1 + 3} \sum_{f=0}^{J_1} {\cal O}_{sin, \,
I}^{(f,{J}_1|\ldots)} \, , \\ \tilde {\cal O}_{sin, \,
II}^{({J}_0,{J}_1|\ldots)} & = & {\cal O}_{sin, \,
II}^{({J}_0,{J}_1|\ldots)} - \frac{1}{2({J}_1-{J}_0+2)}
\sum_{f=0}^{{J}_1-{J}_0-1} {\cal O}_{sin, \, I}^{(f,{J}_1|\ldots)}
\nonumber \\ & & \qquad \qquad \quad \quad \; -
\frac{1}{2({J}_0+2)} \sum_{f=0}^{{J}_0-1} {\cal O}_{sin, \,
I}^{(f,{J}_0 - 1|\ldots)} \nonumber \, .
\end{eqnarray}
Under $(\bar \delta^1)^2$ these operators behave like
\begin{eqnarray}
\tilde {\cal O}_{sin, \, I}^{({J}_0,{J}_1|\ldots)} & \rightarrow &
- 2 \Pi_Z ((\phi_2 Z^{{J}_0+1} \phi_3 Z^{{J}_1-{J}_0}) - (\phi_3
Z^{{J}_0+1} \phi_2 Z^{{J}_1-{J}_0})) \, ,
\\
\tilde {\cal O}_{sin, \, II}^{({J}_0,{J}_1|\ldots)} & \rightarrow
& 0 \, .
\end{eqnarray}
Hence all antisymmetric type I operators of charge $J+1$ are
descendants of the $\tilde {\cal O}_{sin, \, I}$. Further, one may
check from (\ref{app14}),(\ref{app15}) and (\ref{app16}) that the
redefined type II operators are one-loop protected and that the
one-loop mixing matrix of the new type I operators is
\begin{eqnarray}
& \langle \tilde {\cal O}_{sin, \, I}^{({J}_0,{J}_1|\ldots)} \;
\tilde {\cal O}_{sin, \, I}^{\dagger (\bar{J}_0,\bar{J}_1|\ldots)}
\rangle_{g^2} \, = \\ & 8 \, \Pi_Z \Pi_{\bar{Z}} \bigl(
(Z^{{J}_0+1} {\bar{Z}}^{\bar{J}_1-\bar{J}_0})(Z^{{J}_1-{J}_0}
{\bar{Z}}^{\bar{J}_0+1}) - (Z^{{J}_0+1}
{\bar{Z}}^{\bar{J}_0+1})(Z^{{J}_1-{J}_0}
{\bar{Z}}^{\bar{J}_1-\bar{J}_0}) \bigr) \nonumber \, ,
\end{eqnarray}
i.e. the negative of the tree-level mixing of their descendants.

\subsection{Summary}

We have confirmed the intuition that unprotected operators are
essentially type I objects: in the singlet type I
orthogonalised w.r.t. protected operators, in the antisymmetric
representation exactly type I and in the symmetric differences of
type I operators. Amongst the unprotected operators,
the symmetric ones are descendants of antisymmetric operators,
and the antisymmetric operators are descendants of singlets.
This generalises exactly the structure observed in the $J=2$ example
discussed in \cite{op1}.

Further, the one-loop mixing of the singlet operators is
proportional to the tree-level mixing of their symmetric
descendants from which it follows by unitarity that there is no
one-loop protected singlet operator other than the ones we
identified. Likewise, the one-loop mixing of the antisymmetric
type I operators is
\begin{equation}
\langle \Pi_Z \bigl((\phi_2 Z^{J_0} \phi_3 Z^{J_1-J_0}) - (J_0
\leftrightarrow J_1-J_0)\bigr) \Pi_{\bar Z} \bigl((\bar \phi^2
\bar Z^{\bar J_0} \bar \phi^3 \bar Z^{\bar J_1-\bar J_0}) - (\bar
J_0 \leftrightarrow \bar J_1-\bar J_0) \bigr) \rangle_{g^2}
\nonumber
\end{equation}
\vskip -1.1 cm
\begin{eqnarray}
= 4 \, \Pi_Z \Pi_{\bar Z} \bigl( & \; \; \,(Z^{J_0} \bar Z^{\bar
J_0})(Z^{J_1-J_0+1} \bar Z^{\bar J_1-\bar J_0+1}) + (Z^{J_0+1}
\bar Z^{\bar J_0+1})(Z^{J_1-J_0} \bar Z^{\bar J_1-\bar J_0})
\label{relass}
\\ & - (Z^{J_0+1} \bar Z^{\bar J_0})(Z^{J_1-J_0} \bar Z^{\bar J_1-\bar
J_0+1}) - (Z^{J_0} \bar Z^{\bar J_0+1}) (Z^{J_1-J_0+1} \bar
Z^{\bar J_1-\bar J_0})
 \nonumber \\ & - (\bar J_0 \leftrightarrow \bar J_1-\bar J_0) \bigr) \nonumber
\, ,
\end{eqnarray}
which equals the negative of the tree-level mixing matrix of their
symmetric descendants under the ${\cal N} = 4$ supersymmetry
variation $(\delta_4)^2$, see Appendix (\ref{appnot}).

The diagonalisation of a set of operators in ${\cal N} = 4$ SYM at
lowest order in the gauge coupling constant requires
orthogonalisation of the tree-level and the one-loop mixing
matrices \cite{op1,op2}. The change of basis has to coincide
in the three representations as they are connected by
supersymmetry. Let ${\cal V}_{sin}, \, {\cal V}_{as}, \, {\cal
V}_{st}$ denote the tree-level mixing of the singlets, the
antisymmetric and the symmetric BMN operators, respectively.
Further we define ${\cal Z}$ to be the change of basis which
diagonalises the singlets and $\Gamma_1$ the diagonal matrix of
first anomalous dimensions. By reinstating the $g$ in the
commutator terms in the supersymmetry transformations and putting
in the normalisation factors from (\ref{treex}) and (\ref{loopx})
we obtain:\footnote{This requires scaling (\ref{relass}) up by a
factor of 4 in order to match the operator normalisation in the
chain of descendants singlet/antisymmetric/symmetric
representation.}${}^,$\footnote{We thank Y.S.Stanev for a discussion
leading to equation (\ref{yassen}).}
\begin{eqnarray}
{\cal V}_{sin} & = & \phantom{2} \, {\cal Z}^{-1} \, {\cal
Z}^{\dagger \, -1} \, , \label{yassen}
\\ {\cal V}_{as} & = & 2 \, {\cal Z}^{-1} \, \Gamma_1 \, {\cal
Z}^{\dagger \,-1} \, , \nonumber \\ {\cal V}_{st} & = & 4 \, {\cal
Z}^{-1} \, (\Gamma_1)^2 \, {\cal Z}^{\dagger \, -1} \, . \nonumber
\end{eqnarray}

\section{Superspace two-point functions}
\label{suptwopt}

The analysis in Section (\ref{bmnops}) was largely based on the $SO(4)$
decomposition suitable for the BMN limit. In this section we need
the ${\cal N} = 4$ harmonic superspace of \cite{emerymin,howen4,bps3}
and the shortening conditions of \cite{bps3}, i.e. representation theory
of $SU(4)$. Indeed the two pictures are not so different: at low $J$
one can check explicitly that the tree-level orthogonalisation w.r.t.
the protected operators rearranges the $SO(4)$ singlets into $SU(4)$
representations. We illustrate this by the example of the Konishi
operator and the stress energy tensor multiplet ${\cal Q}_{20}$:
for $J=0$ we have the two $SO(4)$ singlets ${\cal O}_1 =(\phi_a
\bar \phi^a), {\cal O}_2 = (Z \bar{Z})$. The operator ${\cal
Q}_{20} = {\cal O}_1 - 2 \, {\cal O}_2$ is protected. Tree-level
orthogonalisation of ${\cal O}_1$ w.r.t. ${\cal Q}_{20}$ gives
${\cal K}_{1} = {\cal O}_1 - 1/3 \, {\cal O}_{2} = 2/3 \, (\phi_I
\bar \phi^I)$, which is the usual lowest component of the Konishi
operator, barring a normalisation factor. For $J=1,2$ this works
in a strictly analogous way: the operators that can pick up a
one-loop anomalous dimension are in the $[0,J,0]$ representation
of $SU(4)$. They have naive scaling dimension $\Delta = J+2$. In
\cite{beisert} this was elaborated for general $J$ but to leading
order in $N$.

Next, observe that the supersymmetry transformations (\ref{dPhi})
are invariant under multiplication by the $SU(4)$ harmonics
$u_{\hat{A}}^A$ of the ${\cal N}=4$ harmonic superspace of
\cite{bps3}, because the $u$ variables do not transform under
Q-supersymmetry. We associate the supersymmetry transformations
with superspace covariant derivatives as follows:
\begin{equation}
\delta_A \, \leftrightarrow \, D^A \qquad
\bar \delta^A \, \leftrightarrow \, \bar D_A
\end{equation}
We will use the double derivative $(\bar D_1)^2$ to go from the
singlets (or rather the $[0,J,0]$ ground states) to the
antisymmetric descendants and the double derivative $(D^4)^2$ to
pass to the symmetric descendants. Note that there can be no field
redefinitions due to the $(\theta \sigma^\mu)
\partial_\mu$ (or c.c.) part of the covariant derivatives, since $\{ \bar D_1,
\, D^4 \} \, = \, 0$ and we put $\theta, \, \bar \theta \, = \, 0$
after differentiation.

According to \cite{bps3} the field content of the highest weight
state of the $[0,J,0]$ multiplets with scaling weight $\Delta$
is correctly reproduced by the product
\begin{equation}
{\cal O}_{[0,J,0]}^{\Delta} \, = \, W_{12}^J (\Psi \bar \Psi)^{(\Delta
- J)/2} \, , \label{classif}
\end{equation}
where $W_{12}$ is the ${\cal N}=4$ on-shell YM multiplet and
$\Psi$ is the ${\cal N}=4$ chiral multiplet. If and only if the operator
has exactly $\Delta = J+2$, the multiplet will obey the shortening
conditions
\begin{eqnarray}
(\bar D^2)_{(rs)} \, {\cal O}_{[0,J,0]}^{J+2} & = & 0 \, , \qquad r,s
\, \epsilon \, \{1,2\} \, , \\ (D^2)^{(\dot r \dot s)} \,
{\cal O}_{[0,J,0]}^{J+2} & = & 0 \, , \qquad \dot r, \dot s \, \epsilon \,
\{3,4\} \, . \nonumber
\end{eqnarray}
The multiplet is ``semishort" in this case. An example is the operator
${\cal D}_{20}$ from \cite{d20,kon} at $J=2$. In Section (\ref{bmnops}) we
had indeed concluded that the one-loop protected operators had vanishing
descendants in the antisymmetric \textbf{6}, which via (\ref{comp6})
coincides with the shortening conditions of the last equation.

The superspace form of the operators (\ref{classif}) suggests a
similar factorisation of their two-point functions:
\begin{equation}
\langle \, {\cal O}_{[0,J,0]}^{\Delta}(z_1) \, \bar {\cal O}_{[0,J,0]}^{\Delta}(z_2)
\, \rangle \, = \, \langle \, W_{12}(1) \, \bar
W^{12}(2) \, \rangle^J \, \Bigl( \langle \, \Psi(1) \, \bar
\Psi(2) \, \rangle \, \langle \, \bar \Psi(1) \, \Psi(2) \,
\rangle \Bigr)^{(\Delta - J)/2} \label{twop}
\end{equation}
This is indeed the unique superspace two-point function with the
correct transformation properties under the full superconformal
group $SU(2,2|4)$, because the three pieces transform locally at
both ends like the constituents $W,\Psi,\bar \Psi$ and because $Q$
and $S$ supersymmetry fix the complete dependence on the spinor
coordinates. For details of the construction see \cite{esok,kuzteis}.
The most elegant way of constructing such two- or three-point functions
is probably superconformal inversion \cite{ivanov}.

For later use we note that the $\theta, \bar \theta = 0$ component of the
function is \cite{esok}
\begin{equation}
\langle \, {\cal O}(x_1) \, \bar {\cal O}(x_2) \, \rangle|_{\theta, \bar \theta
= 0} \, = \,
\frac{\bigl(u_{[1}^A(1)u^1_A(2) u_{2]}^B(1) u^2_B(2)\bigr)^J}
{(x_{12}^2)^\Delta} \, .
\end{equation}
If the harmonics are stripped off, the factor in the numerator will become
the combination of Kronecker deltas typical for a two-point function of
operators in the $[0,J,0]$ irrep of $SU(4)$.

In order to go to the descendant two-point function we apply the
differential operator $(\bar D_1)^2|_1 \, (D^1)^2|_2$. The field
$W_{12}$ is Grassmann analytic: the operator $\bar D_1$
annihilates it. Likewise, the field $\bar W^{12}$ at the second
point obeys the complex conjugate shortening conditions and is
annihilated by $D^1$. The second factor of the two-point function
is also not seen by the differentiation due to its chirality. Hence
the differential operator goes through to the third factor in
(\ref{twop}):
\begin{eqnarray}
& & (\bar D_1)^2|_1 \, (D^1)^2|_2 \, \langle \, {\cal O}(z_1) \, \bar
{\cal O}(z_2) \, \rangle \\ & &
 = \, \langle \, W_{12}(1) \, \bar W^{12}(2) \, \rangle^J \,
\langle \, \Psi(1) \, \bar \Psi(2) \, \rangle^{(\Delta - J)/2} \,
(\bar D_1)^2|_1 \, (D^1)^2|_2 \langle \, \bar \Psi(1) \, \Psi(2)
\, \rangle^{(\Delta - J)/2} \nonumber
\end{eqnarray}
Now, the antichiral/chiral two-point function is simply
\begin{equation}
\langle \, \bar \Psi(1) \, \Psi(2) \, \rangle^{(\Delta-J)/2} \, =
\, e^{- 2 (\bar \theta^I_1 \sigma^\mu \theta_{I \, 2})
\partial_{x_1}} \,
\Bigl( \frac{1}{(x^{R}_{1}-x^{L}_{2})^2} \Bigr)^{(\Delta-J)/2} \, ,
\label{chirachir}
\end{equation}
where the labels 1,2 indicate the points. In the $x$ difference
there is an antichiral $x^{R}$ at point 1 and a chiral $x^{L}$ at point 2.
The differential operator with the outer thetas put to zero will
now act on the exponential only. It produces a contraction of
harmonics times a box operator. The lowest component of the
descendant two-point function is therefore:
\begin{eqnarray}
\Bigl( (\bar D_1)^2|_1 \, (D^1)^2|_2 \, \langle \, {\cal O}(z_1) \, \bar
{\cal O}(z_2) \, \rangle \Bigr)|_{\theta, \bar \theta = 0} \, =
 \label{twopd} \\
\, c_1 (\Delta - J - 2)(\Delta - J) \frac{\bigl(u_{[1}^A(1)
u^1_A(2) u_{2]}^B(1) u^2_B(2)\bigr)^J (u_{1}^C(1) u^1_C(2))^2}
{(x_{12}^2)^{\Delta+1}} \, , \nonumber
\end{eqnarray}
where the constant $c_1$ is some even power of 2 and depends on
how we scale the covariant derivatives on superspace relative to
the ${\cal N} = 4$ supersymmetry transformations (\ref{dPhi})
acting on the elementary fields.

The harmonic projector indicates that the descendant at point 1
carries the $SU(4)$ irrep $[2,J,0]$. We may verify this also
directly from the fact that the all-scalar descendants are type I
operators: the raising operators of $SU(4)$ may be chosen such
that $\phi_3 = \phi_{14} \, \rightarrow \, \phi_2 = \phi_{13} \,
\rightarrow \, Z = \phi_{12}$. These operations annihilate the
antisymmetric type I operators with impurities $\{ \phi_2, \, \phi_3 \}$,
whereby the latter are highest weight states. If we put all boxes
with label 1 into the first row of a Young tableau, all boxes with
label 2 in the second row etc. we see that there is a complete
column, which is to be deleted. The Dynkin labels of the
representation are then once again found to be $[2,J,0]$.

The central observation of this section is another one, though:
while the two-point function of the descendants has the expected
$x$-dependence, it comes with a factor
\begin{equation}
(\Delta - J)(\Delta - J - 2) \, = \, 2 \Bigl( \gamma_1
\frac{g^2}{4 \pi^2} + \Bigl( \frac{\gamma_1^2}{2} + \gamma_2\Bigr)
\frac{g^4}{(4 \pi^2)^2} + \ldots \Bigr) \, . \label{coolf}
\end{equation}
Hence in an orthogonal basis the descendant two-point function has
to lowest order an additional factor $2 \gamma_1 g^2/(4 \pi^2)$ relative to
the ground state two-point function. This explains our observation
of the equality of the matrix of one-loop two-point functions of
the singlets with the tree-level two-point functions of the
antisymmetric descendants. More is true: the second correction to
the anomalous dimension $\gamma_2$ is contained in the constant at
order $g^4$. In the following we will calculate this number for the
$J=0,1,2$ singlet operators for gauge group $SU(N)$.

To go from the antisymmetric operators to the symmetric operators
we use the differential operator $(D^4)^2|_1 \, (\bar D_4)^2|_2$.
By chirality and Grassmann analyticity this acts only on the
second factor of (\ref{twop}). Once again, we obtain a harmonic
projector and a box operator. The $\theta, \bar \theta = 0$
component of this descendant two-point correlator looks like
\begin{eqnarray}
&& \Bigl( (D^4)^2|_1 (\bar D_1)^2|_1 \, (\bar D_4)^2|_2 (D^1)^2|_2
\, \langle \, O(z_1) \, \bar O(z_2) \, \rangle \Bigr)|_{\theta,
\bar \theta = 0}  = \, c_1^2 (\Delta - J - 2)^2    \nonumber \\ &&
\phantom{WW} (\Delta - J)^2 \, \frac {\bigl(u_{[1}^A(1) u^1_A(2)
u_{2]}^B(1) u^2_B(2)\bigr)^J (u_{1}^C(1) u^1_C(2))^2 (u_{C}^4(1)
u^C_4(2))^2} {(x_{12}^2)^{\Delta+2}} \, . \label{twopdd}
\end{eqnarray}
To read off the representation of the descendant at point 1 we
dualise the upper 4 into a lower antisymmetrised $\{1,2,3\}$. On
putting equal labels into the rows of a Young tableau, we find
the Dynkin labels $[2,J,2]$. On the other hand the operators able
to carry one-loop anomalous dimension were differences of type I
objects. The sequence $(D^4)^2 (\bar D_1)^2$ takes the singlets to
symmetric operators with two $\phi_2 = \phi_{13}$ impurities.
Differences of these are indeed annihilated by the raising
operations defined above. Determining the representation from its
highest weight we again fall upon $[2,J,2]$.

Last, the occurrence of the second factor (\ref{coolf}) in
(\ref{twopdd}) explains the equality of the one-loop matrix of
two-point functions of the antisymmetric operators with the
tree-level two-point correlators of their symmetric descendants.

\section{Operators with fermion impurities}
\label{fermimp}

So far we have only considered operators constructed from scalar
fields. In the $[0,J,0]$ irrep of $SU(4)$ at naive scaling weight
$\Delta = J + 2$ there are also the operators (we give the highest
weight state)
\begin{eqnarray}
{\cal Y}_{[0,J,0]} & = & Z^{(J-1)} \psi_{[1} \psi_{2]} \\ \tilde
{\cal Y}_{[0,J,0]} & = & Z^{(J-1)} \bar \psi^{[3} \bar \psi^{4]}
\\ {\cal X}_{[0,J,0]} & = & Z^{(J-2)} (D^\mu Z)(D_\mu Z)
\end{eqnarray}
with some arrangement of the elementary fields in traces of the
gauge group generators.

Likewise, in the $[2,J,0]$ irrep with naive scaling weight $\Delta
= J + 3$ we can write
\begin{equation}
{\cal Y}_{[2,J,0]} \, = \, Z^J \psi_1 \psi_1 \label{yukas} \, .
\end{equation}
In the $[2,J,2]$ with classical dimension $\Delta = J + 4$ there
is no Lorentz scalar involving fermion impurities. These are the
``pure operators'' of \cite{dilat}.

There should be mixing of the all-scalar operators with their two
fermion counterparts. The double-fermion operators have one
elementary field less than their all-scalar partners. On general
field theory grounds we expect mixing like
\begin{eqnarray}
&& {\cal O}_{sin} \, + \, g \, {\cal Y}_{[0,J,0]} \, + \, g \,
\tilde {\cal Y}_{[0,J,0]} \, + \, g^2 \, {\cal O}_{sin} \, + \,
g^2 \, {\cal X}_{[0,2,0]} \, + \, \ldots \, ,
\\ && {\cal Y}_{[0,J,0]} \, + \, g \, {\cal O}_{sin} \, + \, g \, {\cal X}_{[0,J,0]} \, + \, \ldots \, ,
\nonumber \\ && \tilde {\cal Y}_{[0,J,0]} \, + \, g \, {\cal
O}_{sin} \, + \, g \, {\cal X}_{[0,J,0]} \, + \, \ldots \nonumber
\end{eqnarray}
and similarly in the $[2,J,0]$ irrep.

Consider the operators ${\cal Y}_{[2,J,0]}$. We can again distinguish
type I and II operators where the fermions play the role of the
impurities:
\begin{eqnarray}
{\cal Y}_{[2,J,0], \, I} & = & \Pi_Z \, (\psi_1^\alpha Z^{J_0}
\psi_{1 \alpha} Z^{(J_{1}-J_{0})}) \,
, \\ {\cal Y}_{[2,J,0], \, II} & = & \Pi_Z \, (\psi_1^\alpha Z^{J_0})
(\psi_{1 \alpha} Z^{(J_{1}-J_{0})}) \, . \nonumber 
\end{eqnarray}
These operators have the descendants
\begin{eqnarray}
(D^4)^2 \, {\cal Y}_{[2,J,0], \, I} & = & 4 g^2 \, \Pi_Z \, ([Z,\phi_2]
Z^{J_0}
[Z,\phi_2] Z^{(J_{1}-J_{0})}) \, , \\ (D^4)^2 \, {\cal Y}_{[2,J,0], \, II} &
= & 0 \nonumber \, .
\end{eqnarray}
Hence the descendants of the type I double-fermion terms are
differences of type I symmetric operators, quite like the
descendants of the antisymmetric all-scalar operators. We now have
more operators in the $[2,J,0]$ irrep than descendants in the
$[2,J,2]$. Operators with equal descendants must be part of the
same multiplets. Moreover, they have the same naive scaling weight
and $SU(4)$ and spin assignments so that they compete for the same
slot in the multiplets, in a pictorial manner of speaking.

Recall that the derivative of the all-scalar operators comes with
a single power of $g$. We can therefore avoid the problem by
defining the order $g$ all-scalar addition to the ${\cal Y}_{[2,J,0]}$
such that the overall descendant is zero:
\begin{eqnarray}
\hat{\cal Y}_{[2,J,0], \, I}^{(J_{0},J_{1}|\underline{J})} & = &
{\cal Y}_{[2,J,0] \, I}^{(J_{0},J_{1}|\underline{J})} + 2 g (
{\cal O}_{as, \, I}^{(J_{0}+1,J_{1}+1|\underline{J})} \,
 \, - \, {\cal O}_{as, \, I}^{(J_{0},J_{1}+1|\underline{J})} )
\label{yhat}
\\ \hat {\cal Y}_{[2,J,0], \, II}^{(J_{0},J_{1}|\underline{J})}
& = & {\cal Y}_{[2,J,0], \, II}^{(J_{0},J_{1}|\underline{J})}
\nonumber
\end{eqnarray}
Here $ O_{as,I}^{(J_0,J_1+1|\underline J)} = \Pi_i (Z^{J_i})
(\phi_2 Z^{J_0} \phi_3 Z^{(J_1-J_0+1)}) - (\phi_2 \leftrightarrow
\phi_3)$ and the $J_i, \, i > 1$ are equal for ${\cal Y, \,
O}_{as}$. Order $g$ admixtures of type II operators are not
determined by this criterion.

It would be impossible to cancel the descendants of the all-scalar
antisymmetric type I BMN's: they pick up a single power of the
coupling constant $g$ under the supercovariant derivative, whereas
their order $g$ two-fermion addition goes to $g^3$ times a
symmetric descendant.

Lowest order orthogonalisation of the $\hat {\cal Y}$ w.r.t. the
antisymmetric all-scalar BMN operators surprisingly fixes the
latter's two-fermion admixture:
\begin{eqnarray}
0 & = & \langle \, ({\cal O}_{a, \, I}^{f} + g \, {\cal B}
^{f}{}_{h} \, {\cal Y}^{h}) \, ({\cal Y}^{\dagger \, \bar{f}} + g
\, \bar {\cal A}^{\bar{f}}{}_{\bar{h}} \, {\cal O}_{a, \,
I}^{\dagger \, \bar{h}}) \, \rangle_{g^1} \label{konan} \\ & = & g
\, \bar {\cal A}^{\bar{f}}{}_{\bar{h}} \, \langle \, {\cal O}_{a,
\, I}^{f} \, {\cal O}_{a, \, I}^{\dagger \, \bar{h}} \,
\rangle_{g^0} \, + g \, {\cal B}^{f}{}_{h} \, \langle \, {\cal
Y}^{h} \, {\cal Y}^{\dagger \, \bar{f}} \, \rangle_{g^0} \nonumber
\end{eqnarray}
is a non-singular\footnote{Unitarity of the theory requires the
matrix of two-point functions of the ${\cal Y}_{[2,J,0]}$ to be
non-singular at tree-level.} linear system for the matrix ${\cal
B}$ in terms of ${\cal A}$, which is in turn defined by
(\ref{yhat}). Note that the equations are not altered by
admixtures to the $\hat {\cal Y}$ of type II all-scalar operators;
the latter are tree-orthogonal to type I, see Section
(\ref{bmnops}).

The ``anomaly coefficients'' ${\cal B}$ found in the worked examples
below are of very simple form. This matrix can perhaps be found in
closed form by the techniques of Section (\ref{bmnops}).

\section{Operator mixing and anomalous dimensions for J=0,1,2}
\label{opmix}

In this section we restrict to gauge group $SU(N)$ with the obvious
motivation of reducing the number of operators. We wish to
achieve an explicit solution of the mixing, not relying on any
particular limit.

The case $J = 2$ is the most interesting of the three since there
is non-trivial operator mixing, which we will fix up to $g^2$ for
the antisymmetric and the symmetric BMN operators. The discussion of the
fermion mixing in the singlet is postponed to Section (\ref{singsec}).

\subsection{On the singlets}
We have the six $SO(4)$ all-scalar singlets
\begin{eqnarray}
{\cal O}_{sin,1} & = & 2 (Z Z) (\phi_r \bar \phi^r) \, ,
\label{sbas} \\ {\cal O}_{sin,2} & = & (\phi_r \bar \phi^r Z Z) +
(\bar \phi^r \phi_r Z Z) \, , \nonumber \\ {\cal O}_{sin,3} & = &
2 (\phi_r Z \bar \phi^r Z) \, , \nonumber \\ {\cal O}_{sin,4} & =
& 2 (\phi_r Z) (\bar \phi^r Z) \, , \nonumber \\ {\cal O}_{sin,5}
& = & 4 (\phi_r Z) (\bar \phi^r Z) + 2 (Z Z) (\phi_r \bar \phi^r)
- 4 (Z Z) (\bar{Z} Z) \, , \nonumber \\ {\cal O}_{sin,6} & = & 2
(\phi_r Z \bar \phi^r Z) + 2 \Bigl( (\phi_r \bar \phi^r Z Z) +
(\bar \phi^r \phi_r Z Z) \Bigr) - 4 (\bar{Z} Z Z Z) \, , \nonumber
\end{eqnarray}
with $r \, \epsilon \, \{2,3\}$. The last two operators can be
obtained from the 1/2 BPS highest weights $(ZZ)(ZZ)$ and $(ZZZZ)$,
respectively, via the $SU(4)$ lowering operator $\partial_4^{[1}
\partial_3^{2]}$.\footnote{The derivative $\partial_I^J$ replaces a
lower $J$ by a lower $I$.} They are protected. The fourth operator
is of type II and is also (one-loop) protected. The matrix
\begin{equation}
{\cal R} \, = \, \begin{pmatrix} 1 & 0 & 0 & \frac{2 (N^2 - 4)}{3
N^2 - 2} & - \frac{N^2 - 2}{3 N^2 - 2} & 0 \\ 0 & 1 & 0 & -
\frac{N^2 - 4}{N (3 N^2 - 2)} & \frac{N^2 - 4}{5 N (3 N^2 - 2)} &
- \frac{1}{5} \\ 0 & 0 & 1 & - \frac{2}{N} & \frac{2}{5 N} & -
\frac{1}{5} \\ 0 & 0 & 0 & 1 & -\frac{1}{5} & 0 \\ 0 & 0 & 0 & 0 &
1 & 0 \\ 0 & 0 & 0 & 0 & 0 & 1 \end{pmatrix} \label{mbot}
\end{equation}
orthogonalises the first three operators w.r.t. all the protected
operators and the fourth one w.r.t. the 1/2 BPS components. The
first four operators now contain the full $SU(4)$ traces $\phi_a
\bar \phi^a + Z \bar{Z}$, so that they are seen to belong to the
$[0,2,0]$ of $SU(4)$. A second transformation
\begin{equation}
{\cal M} \, =  \, \begin{pmatrix} \frac{3}{4} & 0 & 0 & 0 \\
\frac{N}{2 (N^2 - 2)} & - \frac{1}{2} & \frac{1}{2} & 0 \\ -
\frac{5 (2 N^2 - 3)}{4 N (N^2 + 1)} & \frac{5}{2} & \frac{5}{4} &
0 \\ 0 & 0 & 0 & \frac{5}{6} \end{pmatrix} \label{mmix}
\end{equation}
rotates to the basis of \cite{op1}. The fourth operator is the
semishort ${\cal D}_{20}$.

In \cite{op1} the diagonalisation of the first three operators
was worked out explicitly. Let us define
\begin{equation}
\tilde {\cal O}^f_{sin} \, = \, {\cal R}^f_h \, {\cal O}^h_{sin}
\, .
\end{equation}
The tree-level and the one-loop logarithm of the two-point
functions of the first three operators are diagonalised by the
further transformation
\begin{equation}
{\cal Z } \, = \, {\cal V} \, {\cal M}
\end{equation}
(with the appropriate restriction of ${\cal M}$) and
\begin{equation}
{\cal V} \, = \, \begin{pmatrix} m_1 & 1 & l_1 \\ m_2 & 1 & l_2
\\ m_3 & 1 & l_3 \end{pmatrix} \, , \label{mdiag}
\end{equation}
where
\begin{eqnarray}
m_f & = & \frac{\xi_f}{(N^2+1)(N^2-2)} \, , \\ l_f & = & -
\frac{24}{5} \, m_f^2 \, \frac{(N^2+1)^3}{(3N^2-2)(N^2-4)(N^2-9)}
+ \nonumber \\ & & \phantom{-} \frac{3}{5} \, m_f \,
 \frac{(N^4+4)(N^4-19N^2+10)(N^2+1)}{N
 (3N^2-2)(N^2-2)(N^2-4)(N^2-9)} + \frac{4}{5} \, \frac{N^2 (N^2+1)^2}
 {(N^2-2)^2 (N^2-9)}
 \nonumber
\end{eqnarray}
and the $\xi_f$ are the three roots of the polynomial equation
\begin{eqnarray}
0 & = & 8 N x^3 + (-N^6 + 2 N^4 + 68 N^2 - 40) x^2 +
      (-3 N^9 - 16 N^7 + 132 N^5 - 80 N^3) x \nonumber \\
      & & + (9 N^{12} - 84 N^{10} + 244 N^8 - 224 N^6 + 64 N^4) \, .
\end{eqnarray}
Let
\begin{equation}
\check {\cal O}^f_{sin} \, = \, {\cal Z}^f_h \, \tilde {\cal
O}^h_{sin} \, .
\end{equation}
We find
\begin{eqnarray}
\langle \, {\check {\cal O}}^f_{sin} \, \check {\cal O}^{\dagger
\, f}_{sin} \, \rangle_{g^0} \, = \, \frac{9 (N^2-1)}{(N^2-9)
(N^2-2)^2 (3 N^2-2)} \, \Bigl[ - 6 (N^2 - 6) \xi_f^2 +
\label{treepol}
\\ N (N^6 - 23 N^4 + 112 N^2 - 44) \xi_f + 2 N^2 (N^2 - 4) (3 N^2
- 2) (N^4 - 8 N^2 + 6)\, \Bigr] \nonumber
\end{eqnarray}
and
\begin{eqnarray}
\gamma_{1,f} \, = \, \frac{1}{N^2 (N^2-4) (N^2-2) (3 N^2 - 2)} \,
\Bigl[ 8 N \xi_f^2 - \label{gamma1} \\ (N^6 - 2 N^4 - 68 N^2 + 40)
\xi_f + 2 N^3 (N^2-4) (N^2-6) (3 N^2-2) \, \Bigr] \, . \nonumber
\end{eqnarray}

\subsection{The antisymmetric and symmetric representations}

The classical supersymmetry variation (\ref{dPhi}) of the singlets
(\ref{sbas}) yields
\begin{eqnarray}
{\cal O}_{as, 1} & = & 4 (Z Z) \Bigl( (\phi_2 \phi_3 Z) - (\phi_3
\phi_2 Z) \Bigr) \, , \label{abas} \\ {\cal O}_{as,2} & = & 2
\Bigl( (\phi_2 \phi_3 Z Z Z) - (\phi_3 \phi_2 Z Z Z) \Bigr) - 2
\Bigl( (\phi_2 Z \phi_3 Z Z) - (\phi_3 Z \phi_2 Z Z) \Bigr) \, ,
\nonumber \\ {\cal O}_{as,3} & = & 4 \Bigl( (\phi_2 Z \phi_3 Z Z)
- ( \phi_3 Z \phi_2 Z Z) \Bigr) \, , \nonumber
\\ {\cal O}_{as,4} & = & 0 \, , \nonumber \\ {\cal O}_{as,5} & = & 0 \, ,
\nonumber \\ {\cal O}_{as,6} & = & 0 \, . \nonumber
\end{eqnarray}
Multiplication with ${\cal R}$ acts as the identity because the
protected operators have vanishing descendants. The transformation
to the orthogonal basis is therefore simply ${\cal Z}$.

Next, we have the double-fermion operators
\begin{eqnarray}
{\cal Y}_1 & = & (\psi_1^\alpha \psi_{1 \alpha} Z Z) \, , \label{asys} \\
{\cal Y}_2 & = & (\psi_1^\alpha Z \psi_{1 \alpha}
Z) \, , \nonumber \\ {\cal Y}_3 & = & (\psi_1^\alpha \psi_{1 \alpha})(Z Z) \,
, \nonumber
\\ {\cal Y}_4 & = & (\psi_1^\alpha Z) (\psi_{1 \alpha} Z) \, , \nonumber
\end{eqnarray}
from which we define hatted operators without symmetric
descendants:
\begin{equation}
\hat {\cal Y}^f \, = \, {\cal Y}^f + g \, {\cal A}^f_h \, {\cal
O}^h_{as}
\end{equation} with the
$4 \times 3$ matrix
\begin{equation}
{\cal A} \, = \, \begin{pmatrix} 0 & 1 & 0 \\ 0 & 0 & 1 \\ 1
& 0 & 0 \\ 0 & 0 & 0 \end{pmatrix} \, .
\end{equation}
We can now use equation (\ref{konan}) to determine the two-fermion
term behind the ${\cal O}_{as}$. We define a $3 \times 4$ matrix
of ``anomaly coefficients'':
\begin{equation}
\hat {\cal O}^f_{as} \, = {\cal O}^f_{as} + \frac{g}{4 \pi^2} \,
{\cal B}^f_h \, {\cal Y}^h
\end{equation}
Before we give the actual solution let us switch a comment on the
fermion propagator. From (\ref{dPhi}) we see that $\{Z,\psi_2\}$
can be put into an ${\cal N} = 1$ chiral multiplet:
\begin{equation}
\Phi_1 \, = \, Z(x_L) - \theta^\alpha \, \psi_{2 \alpha}(x_L) +
\ldots
\end{equation}
The ${\cal N} = 1$ directions are $\delta_1, \bar \delta^1$. On
expanding $\langle \, \bar \Phi^1 \, \Phi_1 \, \rangle$ according
to (\ref{chirachir}) we find for the fermion propagator:
\begin{equation}
\langle \, \bar \psi^{2 \dot \alpha}(1) \, \psi_2^\alpha(2) \,
\rangle \, = \, 2 \, (\sigma^\mu)^{\dot \alpha \alpha} \,
\partial_{1 \mu} \, \frac{1}{4 \pi^2 x_{12}^2}
\end{equation}
We can now proceed to solving (\ref{konan}):
\begin{equation}
{\cal B} \, = \, \frac{1}{4} \, \begin{pmatrix} -4 & 4 & -2 N & 0
\\ - 2 N & N & - 1 & 0 \\ 2 N & - 2 N & 0 & 2
\end{pmatrix} \, . \label{bmat}
\end{equation}

The hatted operators are not diagonal to order $g^2$ after the
change of basis by ${\cal Z}$: the tree-level and simple log parts
remain diagonal, of course. But the tree-level two-point functions
of the two-fermion part contribute a non-diagonal mixing at order
$g^2$. In order to cancel this we introduce a $g^2$ addition into
the hatted operators:
\begin{equation}
\hat {\cal O}^f_{as} \, = {\cal O}^f_{as} + \frac{g}{4 \pi^2} \,
{\cal B}^f_h \, {\cal Y}^h + \frac{g^2}{4 \pi^2} \, {\cal C}^f_h
\, {\cal O}^h_{as}
\end{equation}
The matrix ${\cal C}$ can be uniquely determined by the following
three criteria:
\begin{itemize}
\item After changing basis via ${\cal Z}$ we do not wish ${\cal C}$ to
have diagonal components, because these correspond to trivial
rescalings of the operators. In more mathematical terms:
\begin{equation}
\Bigl[ {\cal Z} \, {\cal C} \, {\cal Z}^{-1} \Bigr]_{ff} \, = \, 0
\, , \qquad \forall \, \, f \quad ({\rm no \ sum }) \, .
\end{equation}
\item After the change of basis we want the operators to be orthogonal.
Define ${\cal H}_{f\bar{f}} \, = \, \langle \, {\cal Y}_f \, {\cal
Y}^\dagger_{\bar{f}} \, \rangle_{g^0}$ and ${\cal G}_{h \bar h} \,
= \, \langle \, {\cal O}_{as,h} \, \bar {\cal O}^\dagger_{as, \bar
h} \, \rangle_{g^0}$. The $g^2$ constant contribution is
\begin{equation}
{\cal S} \, = \, \frac{g^2}{4 \pi^2} \, {\cal Z} \, \Bigl(
\frac{1}{4 \pi^2} \, {\cal B} \, {\cal H} \, {\cal B}^\dagger +
{\cal C} \, {\cal G} + {\cal G} \, {\cal C}^\dagger \Bigr) \,
{\cal Z}^\dagger \label{what1}
\end{equation}
and we impose
\begin{equation}
{\cal S}_{fh} \, = \, 0 \, , \qquad f \neq h \, .
\end{equation}
\item Third, we want the $(D^4)^2$ descendants of the operators to stay
orthogonal. Let
\begin{eqnarray}
{\cal O}_{st,f} & = & (D^4)^2 \, {\cal O}_{as,f} \, , \\ {\cal
P}_{f\bar{f}} & = & \langle \, {\cal O}_{st,f} \, {\cal O}^\dagger
_{st,\bar f} \, \rangle_{g^2} \, . \nonumber
\end{eqnarray}
(The descendant correlator has lowest order $g^2$.) The $g^2$
subleading constant contribution to the two-point functions of the
descendants is
\begin{equation}
{\cal T} \, = \, \frac{g^2}{4 \pi^2} \, {\cal Z} \, \Bigl( (-
{\cal B} \, {\cal A} + {\cal C}) \, {\cal P} + {\cal P} ({\cal
C}^\dagger - {\cal A}^\dagger \, {\cal B}^\dagger) \Bigr) \, {\cal
Z}^\dagger \, .
\end{equation}
Orthogonality means
\begin{equation}
{\cal T}_{fh} \, = \, 0 \qquad f \neq h \, .
\end{equation}
\end{itemize}
Note that ${\cal C}$ does not contribute to the diagonal elements of
${\cal S,T}$ because we required it to have vanishing diagonal in the
orthogonal basis.

All in all we have nine linear equations for the nine elements
of the matrix ${\cal C}$. In the case at hand the solution is unique if
somewhat complicated:
\begin{equation}
{\cal C} \, = \, \frac{1}{40 (800 - 1180 N^2 + 116 N^4 + N^6)} \,
\begin{pmatrix} c_{11} & c_{12} & c_{13} \\ c_{21} & c_{22} &
c_{23} \\ c_{31} & c_{32} & c_{33} \end{pmatrix} \, ,
\end{equation}
\begin{eqnarray}
c_{11} & = & -40 N^3 (-24 + 11 N^2) \, , \\ c_{12} & = & 40 (-120 + 212
N^2 - 22 N^4 + 11 N^6) \, ,\nonumber \\ c_{13} & = & -40 (-120 + 152
N^2 - 90 N^4 + 3 N^6) \, , \nonumber \\ c_{21} & = & -5 (280 - 204 N^2
- 206 N^4 + 23 N^6) \, , \nonumber \\ c_{22} & = & -2 N (-920 + 1916
N^2 - 138 N^4 + N^6) \, , \nonumber \\ c_{23} & = & -N (3240 - 2812 N^2
+ 326 N^4 + 3 N^6)) \, , \nonumber \\ c_{31} & = & 10 (-440 + 812 N^2 -
270 N^4 + 13 N^6) \, , \nonumber \\ c_{32} & = & 8 N (-320 + 376 N^2 +
72 N^4 + N^6) \, , \nonumber \\ c_{33} & = & 2 N (-920 + 1436 N^2 + 82
N^4 + N^6) \, . \nonumber
\end{eqnarray}

\newpage

Let us conclude this section with two observations about the
structure of the problem:
\begin{itemize}
\item All equations we have solved here are linear. The only
quadratic problem is the original one of \cite{op1}, i.e. to fix
the $g^0$ mixing. Whereas this is expected for the matrix ${\cal
C}$ it comes as a surprise for ${\cal B}$.
\item In the orthogonal basis the only contribution to the $g^2$
subleading constant comes from the fermion admixture and its
derivative in the antisymmetric representation and the symmetric
representation, respectively. Below we will calculate the second
anomalous dimension from exactly this contribution. This is in
sharp contrast to \cite{forzaitalia}, where fermions were not considered.
\end{itemize}

\subsection{How to extract $\gamma_2$}

Let $\check {\cal O}$ denote the diagonalised operators as before.
We factor the $x$-dependence out of the two-point functions. To
this end we multiply the two-point functions of singlets by
$X_{sin} \, = \, (4 \pi^2 x_{12}^2)^{(J+2)}$, the two-point
functions of their antisymmetric descendants by $x^2_{12} \,
X_{sin}$ and those of the symmetric descendants by $x^4_{12} \,
X_{sin}$. Since we need to keep track of the powers of $g$ in the
supersymmetry transformation from singlet to antisymmetric
representation we also need to scale the ${\cal O}_{as}$ by $g$.
We define therefore $\tilde {\cal S}_{ff} \, = \, (g^2 \, x^2_{12}
\, X_{sin}) \, {\cal S}_{ff}$ and $\tilde {\cal T}_{ff} \, = \, (g^2
\, x^4_{12} \, X_{sin}) {\cal T}_{ff}$. Let the tree-level
normalisation constant of the singlet two-point functions be
denoted as $a_0$.

What remains of the two-point functions in the antisymmetric
representation is:
\begin{eqnarray}
\langle \, \check {\cal O}_{as,f} \, \check {\cal
O}^\dagger_{as,f} \, \rangle & = & 2 \, (a_{0,f} + a_{2,f} \,
\frac{g^2}{4 \pi^2}) \Bigl( \gamma_{1,f} \, \frac{g^2}{4 \pi^2} -
\gamma_{1,f}^2 \, \frac{g^4}{(4 \pi^2)^2} \, (\ln(x^2_{12}) +
\alpha) \Bigr) \nonumber \\ & \phantom{=} & + \, \tilde {\cal
S}_{ff} + O(g^6) \label{astop}
\end{eqnarray}
Similarly:
\begin{eqnarray}
\langle \, \check {\cal O}_{st,f} \, \check {\cal
O}^\dagger_{st,f} \, \rangle & = & 4 \, (a_{0,f} + a_{2,f} \,
\frac{g^2}{4 \pi^2}) \Bigl( \gamma_{1,f}^2 \, \frac{g^4}{(4
\pi^2)^2} - \, \gamma_{1,f}^3 \, \frac{g^6}{(4 \pi^2)^3} \,
(\ln(x^2_{12}) + \alpha) \Bigr) \nonumber \\ & \phantom{=} & +
\tilde {\cal T}_{ff} + O(g^8) \label{symtop}
\end{eqnarray}
Recall that the one-loop graph calculations lead in both cases to
only one type of divergent $x$-space integral, see (\ref{loopx}).
The constant $\alpha$ behind the logarithm is therefore the same
in both correlators for any well-defined regularisation scheme.
The other constant $a_2$ can arise from the derivative of an order
$g$ double-fermion addition $\tilde {\cal Y}_{[0,J,0]}$ to the
singlet operators. After all, hidden in ${\cal T}_{ff}$ the
correct two-point function of the symmetric descendants contains a
rescaling stemming from the derivative of the anomaly of the
antisymmetric operators. Even if generally $a_2 \neq 0$, as an
overall normalisation it is not affected by the differentiation
$(D^4)^2$ which leads from the first to the second correlator.

Next,
\begin{eqnarray}
\tilde {\cal S}_{ff} & = & \frac{g^4}{(4 \pi^2)^2} \, \frac{1}{2}
\, \frac{9 (N^2-1)}{(N^2-9) (N^2-2)^2 (3 N^2-2)}
\\ & & N \, \Bigl( 20 N (N^2-10) \xi_f^2 + (1000 - 2028 N^2 + 498
N^4 - 25 N^6 - N^8) \xi_f + \nonumber \\ & & \phantom{N \, \Bigl(}
N (N^2-4) (3 N^2-2) (84 + 96 N^2 - 35 N^4 + 3 N^6) \Bigr)
\nonumber
\end{eqnarray}
and one finds
\begin{equation}
\tilde {\cal T}_{ff} \, = \, 4 \gamma_{1,f} \frac{g^2}{4 \pi^2} \;
\tilde {\cal S}_{ff} \label{tiss} \, .
\end{equation}
The difference (\ref{symtop}) $-$ $2 \gamma_{1,f} g^2/(4 \pi^2)$
(\ref{astop}) is independent of $a_{2,f},\alpha$.

We want to match it with the corresponding difference of abstract
superspace functions (\ref{twopdd}) $-$ $2 \gamma_{1,f} g^2/(4
\pi^2)$ (\ref{twopd}). The arbitrary normalisation $c_1$ in those
equations can be fixed by individually matching the lowest order
of (\ref{twopd}) and (\ref{astop}). We also take into account the
tree-level normalisation $a_{0,f}$. The resulting equation is
\begin{equation}
\tilde {\cal S}_{ff} \, = \, a_{0,f} \, (\gamma_{1,f}^2 + 2
\gamma_{2,f}) \, \frac{g^4}{(4 \pi^2)^2} \label{master} \, .
\end{equation}
Explicitly:
\begin{eqnarray}
\gamma_{2,f} & = & - \frac{1}{4 N (N^2-4) (N^2-2)^2 (3 N^2-2) (800
- 1180 N^2 + 116 N^4 + N^6)}   \nonumber \\ & & \Bigl[ 8 N (-13680
+ 24976 N^2 - 9016 N^4 + 668 N^6 + 5 N^8) \xi_f^2 + \\ & &
\phantom{ \Bigl[} (547200 - 1995840 N^2 + 2250272 N^4 - 791920 N^6
+ \nonumber \\ & & \phantom{\Bigl[ (} 61640 N^8 + 6068 N^{10} -
634 N^{12} - 5 N^{14}) \xi_f + \nonumber
\\ & & \phantom{\Bigl[} 2 N (N^2-4) (3 N^2-2) (20800 + 4320 N^2 -
72048 N^4 + \nonumber \\ & & \phantom{\Bigl[ 2 N (N^2-4) (3 N^2-2)
(} 42112 N^6 - 6980 N^8 + 338 N^{10} + 3 N^{12}) \Bigr] \nonumber
\end{eqnarray}
(The dimension of the singlets was defined as $\Delta \, = \,
J + 2 + \gamma_1 \, \frac{g^2}{4 \pi^2} + \gamma_2 \,
\frac{g^4}{(4 \pi^2)^2} + \ldots$)

\subsection{$J = 1$}

The $SO(4)$ singlets are
\begin{eqnarray}
{\cal O}_{sin,1} & = & (\phi_a \bar \phi^a Z) + (\bar \phi^a
\phi_a Z) \, , \\ {\cal O}_{sin,2} & = & (\phi_a \bar \phi^a Z) +
(\bar \phi^a \phi_a Z) - 2 (\bar Z Z Z) \, . \nonumber
\end{eqnarray}
The second operator equals $\frac{2}{3} \, \partial_4^{[1} \,
\partial_3^{2]} \, (ZZZ)$, whence it is a component of a 1/2 BPS state.
Tree-level orthogonalisation gives
\begin{equation}
\tilde {\cal O}_{sin} \, = \, {\cal O}_{sin,1} - \frac{1}{2} \,
{\cal O}_{sin,2} \, = \, \frac{1}{2} \, \Bigl( (\phi_I \bar \phi^I
Z) + (\bar \phi^I \phi_I Z) \Bigr) \, , \quad I \, \epsilon \,
\{1,2,3\} \, ,
\end{equation}
which has tree-level two-point function
\begin{equation}
\langle \, \tilde {\cal O}_{sin} \, \tilde {\cal O}^\dagger_{sin}
\, \rangle \, = \, \frac{2 (N^2-1)(N^2-4)}{N}
\end{equation}
and first anomalous dimension
\begin{equation}
\gamma_1 \, = \, 2 N \, .
\end{equation}

In the antisymmetric representation we have
\begin{eqnarray}
{\cal O}_{as} & = & 2 \Bigl( (\phi_2 \phi_3 Z Z) - (\phi_3 \phi_2
Z Z) \Bigr) \, ,
\\ {\cal Y} & = &  (\psi_1^\alpha \psi_{1 \alpha} Z) \, ,
\end{eqnarray}
which receive the order $g$ corrections
\begin{eqnarray}
\hat {\cal Y} & = & {\cal Y} + g \, {\cal O}_{as} \, , \\ \hat
{\cal O}_{as} & = & {\cal O}_{a} + \frac{g}{4 \pi^2} \,
\frac{N}{4} \, {\cal Y} \, .
\end{eqnarray}
The matrix ${\cal C}$ is absent. Repeating the same steps as above we
find
\begin{equation}
\gamma_2 \, = \, - \frac{3 N^2}{2} \, .
\end{equation}

\subsection{$J = 0$}

The $SO(4)$ singlets are
\begin{eqnarray}
{\cal O}_{sin,1} & = & 2 (\phi_a \bar \phi^a) \, ,
\\ {\cal O}_{sin,2} & = & 2 (\phi_a \bar \phi^a)
- 4 (\bar Z Z) \, . \nonumber
\end{eqnarray}
The second operator is a component of the stress-energy tensor
multiplet ${\cal O}_{20}$, as mentioned above. Tree-level
orthogonalisation completes ${\cal O}_{sin,1}$ to the Konishi
operator
\begin{equation}
\tilde {\cal O}_{sin} \, = \, {\cal O}_{sin,1} - \frac{1}{3} \,
{\cal O}_{sin,2} \, = \, \frac{4}{3} (\phi_I \bar \phi^I) \, ,
\end{equation}
with tree-level two-point function
\begin{equation}
\langle \, \tilde {\cal O}_{sin} \, \tilde {\cal O}^\dagger_{sin}
\, \rangle \, = \, \frac{16 (N^2-1)}{3} \, .
\end{equation}
Its first anomalous dimension is
\begin{equation}
\gamma_1 \, = \, 3 N \, .
\end{equation}

In the antisymmetric representation we have
\begin{eqnarray}
{\cal O}_{as} & = & 4 (Z [\phi_2, \phi_3]) \, ,
\\ {\cal Y} & = &  (\psi_1^\alpha \psi_{1 \alpha}) \, ,
\end{eqnarray}
which receive the order $g$ corrections
\begin{eqnarray}
\hat {\cal Y} & = & {\cal Y} + g \, {\cal O}_{as} \, , \\ \hat
{\cal O}_{as} & = & {\cal O}_{as} + \frac{g}{4 \pi^2} \,
\frac{N}{2} \, {\cal Y} \, .
\end{eqnarray}
For the second correction to the anomalous dimension we find
\begin{equation}
\gamma_2 \, = \, - 3 N^2 \, .
\end{equation}

\section{Test against the two-loop dilation operator}
\label{dilat}

According to \cite{dilat} the correctly orthogonalised operators
and their dimensions can be obtained from the eigenvalue problem
of the dilation operator represented as a functional
differentiation on a given operator basis. The two-loop
differential operator adapted to the symmetric two impurity BMN
operators is
\begin{equation}
\triangle \, = \, \triangle_0 + \frac{g^2}{16 \pi^2} \,
\triangle_2 + \frac{g^4}{(16 \pi^2)^2} \, \triangle_4 + \ldots
\end{equation}
with
\begin{eqnarray}
\triangle_0 & = & (Z \check Z) + (\phi_2 \check \phi_2) \, , \\
\triangle_2 & = & - 2 ([\phi_2,Z][\check \phi_2, \check Z]) \, ,
\nonumber \\ \triangle_4 & = & - 2 :([[\phi_2,Z],\check Z][\check
\phi_2, \check Z],Z]): - :([[\phi_2,Z],\check \phi_2][\check
\phi_2,\check Z],\phi_2]): \nonumber \\ & & - 2
\phantom{:}([[\phi_2,Z],T^a][\check \phi_2, \check Z], T^a]) \, .
\nonumber
\end{eqnarray}
In this formula $\check Z = \delta/\delta Z$ etc. and the normal
ordering means that the functional derivatives do not act within
the operator itself.

A first consistency check with our material is the zero eigenspace
of $\triangle_2$ and $\triangle_4$. For $J=0,1,2$ with gauge group
$SU(N)$ we find agreement: the orthogonal complement of the zero
eigenspace is given by differences of type I symmetric operators.

Let us restrict our attention to such operators. The dilation
operators $\triangle_{0,2,4}$ send the space into itself, so that
we may associate matrices with them:
\begin{eqnarray}
\triangle_0 & \cong & (J+4) \, \mathbb{I} \, \\ \triangle_2 &
\cong & D_2 \, , \nonumber \\ \triangle_4 & \cong & D_4 \, .
\nonumber
\end{eqnarray}
Let our operators compose a vector $\underline {\cal O}$. The operators
as well as the eigenvalues (viz the dimensions) have an expansion
in $g^2$ like the dilation operator:
\begin{eqnarray}
\underline {\cal O} & = & {\underline {\cal O}}_0 + \frac{g^2}{4 \pi^2}
 \, {\underline {\cal O}}_2 +
\ldots \, \\ \lambda & = & \lambda_0 + \frac{g^2}{4 \pi^2} \, \lambda_2 +
\frac{g^4}{(4 \pi^2)^2} \, \lambda_4 \, . \nonumber
\end{eqnarray}
We arrange the eigenvalues into a diagonal matrix $\Lambda$. The
eigenvalue problem
\begin{equation}
\triangle \, \underline {\cal O} \, = \, \Lambda \, \underline {\cal O}
\end{equation}
expanded through $g^2$ gives the equation
\begin{equation}
D_2 \, {\underline {\cal O}}_0 \, = \, \Lambda_2 \, {\underline {\cal O}}_0 \, .
\end{equation}
Suppose we solve this by going to a diagonal basis (Above we
denoted this by $\check{\underline {\cal O}}$. For simplicity we do not
change all the symbols.) The next order of the eigenvalue problem
is
\begin{equation}
D_2 \, {\underline {\cal O}}_2 + D_4 \, {\underline {\cal O}}_0 \, = \Lambda_2
\, {\underline {\cal O}}_2 + \Lambda_4 \, {\underline {\cal O}}_0 \, .
\end{equation} As before, we introduce the $g^2$ mixing ${\cal C}'$
\begin{equation}
{\underline {\cal O}}_2 \, = \, {\cal C}' \, {\underline {\cal O}}_0
\end{equation}
to obtain
\begin{equation}
D_4 \, = \, (\Lambda_2 \, {\cal C}' - {\cal C'} \, \Lambda_2) \, +
\Lambda_4 \, .
\end{equation}
In this equation the first term on the r.h.s. has zero diagonal,
because $\Lambda_2$ is diagonal. As a consequence, in the basis
$\check{\underline {\cal O}}$ the diagonal of the matrix $D_4$ contains
the $g^4$ contribution to the eigenvalues and the off-diagonal
part has to be cancelled by the $g^2$ operator admixture defined
by ${\cal C}'$. Note that the matrix ${\cal C}'$ has no influence on
$\Lambda_4$, quite like in our calculation above. If we require the absence of
trivial $g^2$ rescalings, ${\cal C}'$ is once again uniquely determined.

For $J=0,1$ agreement with our method is immediate. For $J=2$ we used
150 digits precision numerics under $Mathematica$ to calculate in the
dilation operator method. For the $g^2$ operator mixing we must have
\begin{equation}
\Bigl[ {\cal Z} ( - {\cal B \, A \, + \, C } ) {\cal Z}^{-1} \, - \,
{\cal C}' \Bigr]_{fh} \, = \, 0 \, , \qquad f \neq h \, ,
\end{equation}
while the values for $\gamma_2$ can be directly compared. To the given
accuracy there are no deviations.

\section{Operator Mixing in the Singlet}
\label{singsec}

Recall that the $SO(4)$ singlet contains the operators ${\cal
O}_{sin,J}, \, {\cal Y}_{[0,J,0]}, \, \tilde {\cal Y}_{[0,J,0]}$
and ${\cal X} = Z^{J-2} (D^\mu Z) (D_\mu Z)$ in some gauge trace
arrangement. We will not dedicate much attention to the ${\cal X}$
objects. Mixing between the all-scalar and two-fermion operators
certainly exists: the author learned in discussion\footnote{We are
grateful to M.Bianchi, G.Rossi and Y.S.Stanev.} that for example
${\cal D}_{20}$ at $J=2$ has subleading corrections.

The framework of this article gives an elegant way of determining
the lowest order terms: as in the antisymmetric representation we
can fix the order $g$ correction to the ${\cal Y}$ by
supersymmetry. One can then like in equation (\ref{konan}) use
lowest order orthogonalisation to establish the leading correction
to the ${\cal O}_{sin}$, too.

For $J=2$ there are only two Yukawa structures \cite{kon}:
\begin{eqnarray}
\hat {\cal K}_{20}^+ \, = \, (Z \psi^\alpha_{[1}
\psi^{\phantom{\alpha}}_{2] \alpha}) + g \, \check {\cal A}_f \,
{\cal O}^f_{sin} + \frac{g N}{32 \pi^2} ((\partial^\mu
Z)(\partial_\mu Z)) + \ldots \label{connery1} \\ {\hat {\cal
K}}_{20}^- \, = \, (Z \psi_{\dot \alpha}^{[3} \psi_{\phantom{\dot
\alpha}}^{4] \dot \alpha}) + g \, \check {\cal A}_f \, {\cal
O}^f_{sin} + \frac{g N}{32 \pi^2} ((\partial^\mu Z)(\partial_\mu
Z)) + \ldots \label{connery2}
\end{eqnarray}
with
\begin{equation}
\underline{\check {\cal A}} \, = \, \{0,-1,1,0,0,0\} \, .
\end{equation}
Both of these are level four descendants of the Konishi scalar
${\cal K}_1$. In the same way, all the operators ${\cal Y}, \tilde
{\cal Y}$ are at least level two descendants: the naive supersymmetry
variation $(\bar \delta^2)^2$ acts on antisymmetric all-scalar BMN
operators with impurities $\{ \phi_2, \phi_3 \}$ like
\begin{eqnarray}
& (\bar D_2)^2 &  \Pi_Z ((\phi_2 Z^p \phi_3 Z^{k-p}) - (p
\leftrightarrow k-p)) \, = \\ & - & \phantom{g} \, \Pi_Z (\bar
\psi^{[3}_{\dot \alpha} Z^p \bar \psi^{4] \dot \alpha} Z^{k-p})
\nonumber \\ & - & g \, \Pi_Z ((\phi_I Z^{p+1} \bar \phi^I
Z^{k-p}) + (\bar \phi^I Z^{p+1} \phi_I Z^{k-p})) \nonumber \\ & +
& g \, \Pi_Z ((\phi_I Z^p \bar \phi^I Z^{k-p+1}) + (\bar \phi^I
Z^p \phi_I Z^{k-p+1}))  \, ,\nonumber
\\ & (\bar D_2)^2 & \Pi_Z ((\phi_2 Z^p)(\phi_3 Z^{k-p}) - (p
\leftrightarrow k-p)) \, = \\ & - & \phantom{g} \, \Pi_Z (\bar
\psi^{[3}_{\dot \alpha} Z^p)(\bar \psi^{4] \dot \alpha} Z^{k-p})
\, . \nonumber
\end{eqnarray}
Up to order $g^2$ there should be no anomaly. The supersymmetry
transformation $(D^3)^2$ acting on ${\cal O}_{as}$ with impurities
$\{ \phi_2, \bar \phi^3 \}$ gives analogous formulae with $\tilde
{\cal Y}$ replaced by ${\cal Y}$ but remarkably an identical
${\cal O}_{sin}$ part.

Similarly, the ${\cal X}$ operators are descendants of the ${\cal
Y}_{[2,J,0]}$:
\begin{eqnarray}
& (\bar D_2)^2 & \Pi_Z (\psi^\alpha_1 Z^p \psi_{1 \alpha} Z^{k-p})
\, = \\ & - & \Pi_Z ((D^\mu Z) Z^p (D_\mu Z) Z^{k-p}) \nonumber
\\ & - & g \Pi_Z (\psi^\alpha_{[1} Z^{p+1} \psi_{2] \alpha} Z^{k-p}) +
g \Pi_Z (\psi^\alpha_{[1} Z^p \psi_{2] \alpha} Z^{k-p+1}) \, ,
\nonumber \\ & (\bar D_2)^2 & \Pi_Z (\psi^\alpha_1 Z^p) (\psi_{1
\alpha} Z^{k-p}) \, = \\ & - & \Pi_Z ((D^\mu Z) Z^p) ((D_\mu Z)
Z^{k-p}) \, . \nonumber
\end{eqnarray}
In Section (\ref{fermimp}) we had defined the operators
$\hat {\cal O}_{as}^f \, = \, {\cal O}_{as}^f + \frac{g}{4 \pi^2} \,
{\cal B}^f_h \, {\cal Y}^h_{[2,J,0]}$. From the formulae above their
descendants under $(\bar D_2)^2$ are of the form
\begin{equation}
(\bar D_2)^2 \, \hat {\cal O}_{as,J-1} \, = \, \tilde {\cal
Y}_{[0,J,0]} + g {\cal O}_{sin,J} + \frac{g}{4 \pi^2} \, Z^{J-2}
(\partial^\mu Z)(\partial_\mu Z) + \ldots \label{yukDesc}
\end{equation}
where the dots denote terms of order $g^2$ and higher.

On the other hand, for the ${\cal O}_{sin}$ we expect mixing like
\begin{equation}
\hat {\cal O}^d_{sin} = {\cal O}^d_{sin} + \frac{g}{4 \pi^2} \,
\check {\cal B}^d_e \, ({\cal Y}^e + \tilde {\cal Y}^e) +
\frac{g^2}{4 \pi^2} \, \check {\cal C}^d_f \, {\cal O}^f_{sin} +
\frac{g^2}{4 \pi^2} \, \check {\cal D}^d_h \, {\cal X}^h \,  +
\ldots
\end{equation}
These operators are apparently primary. They ought to be
orthogonal to the descendants (\ref{yukDesc}) which belong to
multiplets with highest weights at lower $J$. Order $g$
orthogonalisation fixes the matrix $\check {\cal B}$ just as in
eq. (\ref{konan}) in the antisymmetric representation. The
resulting linear system of equations is once again guaranteed to
be non-singular due to the unitarity of the theory. Note that the
coefficients for ${\cal Y}$ and $\tilde {\cal Y}$ always come out
equal.

Let us carry out this programme for the case $J=2$. For the ${\cal
O}_{sin}$ we use the basis (\ref{sbas}). The Yukawa like operators
are (\ref{connery1}), (\ref{connery2}) from above. Order $g$
orthogonality w.r.t. the $\hat {\cal O}_{sin}$ determines
\begin{equation}
\underline{\check {\cal B}} \, = \, \{ 1, \frac{N}{4},
-\frac{N}{2}, -\frac{1}{2}, 0, 0 \} \, .
\end{equation}
The double derivative $(\bar D_1)^2$ replaces both fermions in
$\tilde {\cal Y}$ by commutators of scalars, whereas we find
\begin{equation}
(\bar D_1)^2 \, (Z \psi^\alpha_{[1} \psi^{\phantom{\alpha}}_{2]
\alpha}) \, = \, - g \, (\psi^\alpha_1 [\psi^{\phantom{\alpha}}_{1
\alpha}, Z] Z) \, , \label{varyy}
\end{equation}
because the first transformation sends $\psi_{2 \alpha}$ to a
Yang-Mills covariant derivative $D = \partial + g [A, \cdot]$ on
$Z$ and the second variation converts the vector field in the
derivative into another fermion.

Second, $(\bar D_1)^2$ when acting on ${\cal O}_{sin}$ produces
not only the naive descendant but also the generalised Konishi
anomaly. We repeat formula (\ref{Massimo})
\begin{equation}
{\cal F}_K \, = \, - \frac{1}{16} \, \frac{g^2}{4 \pi^2} \,
\Biggl( \Bigl( \psi^\alpha_1 \Bigl[ \psi^{\phantom{\alpha}}_{1
\alpha}, \frac{\delta}{\delta \phi_I} \Bigr] \frac{\delta}{\delta
\bar \phi^I} \Bigr) + (\phi \leftrightarrow \bar \phi) \Biggr)
\nonumber
\end{equation}
for the anomalous part of the supersymmetry. The generalised
Konishi anomaly does not lead to order $g^3$ all-scalar admixtures
and it annihilates the 1/2 BPS states. Remarkably, the operator
${\cal F}_K$ changes the gauge trace structures whereas the
naive supersymmetry transformations never do so.

Let us define the coefficient matrix ${\cal B}_1$ by
\begin{equation}
{\cal F}_K \, {\cal O}_{sin,f} \, = \, \frac{g^2}{4 \pi^2} \,
{\cal B}^h_{1 f} \, {\cal Y}_h
\end{equation}
with the four ${\cal Y}_{[2,2,0]}$ from equation (\ref{asys}).
Similarly, we arrange the double-fermion terms from the naive
$(\bar D_1)^2$ acting on ${\cal Y}_{[0,2,0]}$ in each $\hat {\cal
O}^f_{sin}$ into a form $\frac{g^2}{4 \pi^2} \, {\cal B}^h_{2 f}
\, {\cal Y}_h$. We find
\begin{equation}
{\cal B}_1 \, + \, {\cal B}_2 \, = \, \frac{1}{4} \begin{pmatrix}
0 & 0 & - 2 N & 0 \\ - N & 0 & - 1 & 0 \\ 0 & 0 & 0 & 2
\\ - 2 & 2 & 0 & 0 \end{pmatrix} \, + \, \frac{1}{4}
\begin{pmatrix} -4 & 4 & 0 & 0 \\ -N & N & 0 & 0 \\ 2N & -
2 N & 0 & 0 \\ 2 & -2 & 0 & 0 \end{pmatrix} \, . \label{beauty}
\end{equation}
The two empty lines relating to ${\cal O}_{sin,5}, {\cal
O}_{sin,6}$ have been omitted. In the fourth line the two
contributions cancel: the absence of the descendant is necessary
for ${\cal D}_{20}$ to be semishort. For the first three operators
the sum of the two matrices exactly reproduces ${\cal B}$ from equation
(\ref{bmat}).

In the cases $J=0,1$ there are no ${\cal Y}, \tilde {\cal Y}$
operators in the singlet (also no ${\cal X}$). Correspondingly,
the generalised Konishi anomaly accounts for the whole double
fermion admixture to the antisymmetric descendants of the long
operators.

Let us proceed by fixing the order $g^2$ additions to ${\cal
D}_{20}$. The two Konishi descendants are $\hat {\cal K}_{20}^+ =
(D^3)^2 (D^4)^2 \, {\cal K}_1$ and $\hat {\cal K}_{20}^- = (\bar
D_1)^2 (\bar D_2)^2 \, {\cal K}_1$. Clearly, the first operator is
annihilated by $\{D^3, D^4\}$. We conclude that to order $g^2$ $\{
(D^3)^2, (D^3 D^4), (D^4)^2 \}$ take its Yukawa part into the
negative of the derivatives of the ${\cal O}_{sin}$ admixture, and
similarly for $\{ (\bar D_1)^2, (\bar D_1 \bar D_2), (\bar D_2)^2
\}$ acting on $\hat {\cal K}_{20}^-$. The sum
\begin{equation}
\hat {\cal O}^4_{sin} \, = \, {\cal O}^4_{sin} - \, \frac{1}{2} \,
\frac{g}{4 \pi^2} \, ((Z \psi^\alpha_{[1}
\psi^{\phantom{\alpha}}_{2] \alpha}) + (Z \psi_{\dot \alpha}^{[3}
\psi_{\phantom{\dot \alpha}}^{4] \dot \alpha})) - \frac{1}{2} \,
\frac{g^2}{4 \pi^2} \, \check {\cal A}_f \, {\cal O}^f_{sin}
\end{equation}
has vanishing descendants (up to order $g^3$) in all six
components of the antisymmetric representation, because under each
of $\{ (\bar D_1)^2, \ldots, (D^3)^2, \ldots \}$ the derivative of
one Yukawa term cancels the generalised Konishi anomaly and that
of the other compensates the variation of the $g^2$ scalar
remixing.

In the basis
\begin{equation}
\tilde {\cal O}^f_{sin} = {\cal R}^f_h \, {\cal O}^h_{sin}
\end{equation}
the fourth operator is ${\cal D}_{20}$. The vector $\check {\cal
A}$ goes into
\begin{equation}
\check {\cal A} \, = \, \{ 0,-1,1,\frac{5N}{3N^2-2},0,0 \} \, .
\end{equation}
Thus ${\cal D}_{20}$ picks up a $g^2$ rescaling which can omit.
The result is:
\begin{equation}
\hat {\cal D}_{20} \, = \, {\cal D}_{20} - \, \frac{1}{2} \,
\frac{g}{4 \pi^2} \, ((Z \psi^\alpha_{[1}
\psi^{\phantom{\alpha}}_{2] \alpha}) + (Z \psi_{\dot \alpha}^{[3}
\psi_{\phantom{\dot \alpha}}^{4] \dot \alpha})) + \frac{1}{2} \,
\frac{g^2}{4 \pi^2} ({\cal O}^2_{sin} - {\cal O}^3_{sin}) \,
\end{equation}
with
\begin{equation}
{\cal D}_{20} \, = \, \frac{2}{5} (3 \, (\phi_I Z)(\bar \phi^I Z)
- (\phi_I \bar \phi^I)(Z Z)) \, .
\end{equation}
An addition of $g^2 \, ((\partial Z)(\partial Z))$ remains
undetermined.

In the new basis let us write (for simplicity we omit the tilde on
all symbols)
\begin{equation}
\hat {\cal O}^f_{sin} \, = \, {\cal O}^f_{sin} + \frac{g}{4 \pi^2}
\, \check {\cal B}^f \, ({\cal Y} + \tilde {\cal Y}) +
\frac{g^2}{4 \pi^2} \, \check {\cal C}^f_h \, {\cal O}^h_{sin} +
\frac{g^2}{4 \pi^2} \, \check {\cal C}^f_4 \, {\cal D}_{20} \, ,
\end{equation}
for $f,h \, \epsilon \, \{1,2,3\}$. The condition $\langle \, \hat
{\cal O}^f_{sin} \, \hat {\cal D}_{20}^\dagger \, \rangle_{g^2} \,
= \, 0$ gives
\begin{equation}
\check {\cal C}_4 \, = \, \{ \frac{10 N (N^2+1)}{(3 N^2-2)^2},
\frac{5}{4} \frac{(3 N^4-8)}{(3 N^2-2)^2}, - \frac{5 (N^2-2)}{2(3
N^2-2)} \} \, .
\end{equation}
For completeness we mention that in this basis
\begin{equation}
\check {\cal B} \, = \, \{\frac{2 (N^2+1)}{3 N^2-2},
\frac{3N^4-8}{4 N (3 N^2-2)}, -\frac{N^2-2}{2 N} \} \, .
\end{equation}

The matrix $\check {\cal C}^f_h$ can be fixed as follows:
\begin{itemize}
\item After changing to the tree- and one-loop logarithm orthogonal
basis it should not have diagonal elements.
\item In the orthogonal basis we want it to cancel $g^2$ off-diagonal
contributions introduced by the Yukawa admixtures. The resulting
equation is like (\ref{what1}) in the antisymmetric
representation. Let $\check {\cal H} \, = \, \langle {\cal Y
\,Y}^\dagger \rangle_{g^0} + \langle \tilde {\cal Y} \, \tilde
{\cal Y}^\dagger \rangle_{g^0}$ and $\check {\cal G}_{h \bar h} \,
= \, \langle {\cal O}_{sin,h} \, {\cal O}^\dagger_{sin,\bar h} \,
\rangle_{g^0}$. The order $g^2$ constant part in the mixing is
\begin{equation}
\check {\cal S} \, = \, \frac{g^2}{4 \pi^2} \, {\cal Z} ( \check
{\cal B} \otimes \check {\cal B}^\dagger   \frac{1}{4 \pi^2}
\check {\cal H} + \check {\cal C} \, \check {\cal G} + \check {\cal
G} \, \check {\cal C}^\dagger ) {\cal Z}^\dagger
\end{equation}
and we demand $\check {\cal S}_{fh} = 0, \; f \neq h$ as before.
\item The descendants in the antisymmetric representation must
be as determined in the preceding sections. (We have already
checked the double-fermion terms.) The $g^2$ subleading all-scalar
contribution in the antisymmetric descendants is
\begin{equation}
(\bar D_1)^2 \, {\cal O}^f_{sin} |_{g^3} \, = \, \frac{g^3}{4
\pi^2} \, (- \check {\cal B} \otimes \check {\cal A}^\dagger +
\check {\cal C})^f_h \, {\cal O}^f_{as} \, ,
\end{equation}
but we have to take care of the fact that the first term in the
bracket actually introduces diagonal rescalings of the
antisymmetric operators in the orthogonal basis; we had banned
these in the above. We can therefore only impose
\begin{equation}
\Bigl({\cal Z} (- \check {\cal B} \otimes \check {\cal A}^\dagger
+ \check {\cal C} \, - \, {\cal C}) {\cal Z}^{-1}\Bigr)_{fh} \, =
\, 0 \, , \qquad f \neq h \, .
\end{equation}
\end{itemize}
These are all in all 12 equations on 9 matrix elements, which
constitutes a stringent consistency check. A solution does indeed
exist:
\begin{equation}
\check {\cal C} \, = \, \frac{1}{20 N (3 N^2 - 2) (800 - 1180 N^2
+ 116 N^4 + N^6)} \, \begin{pmatrix} \check c_{11} & \check c_{12}
& \check c_{13} \\ \check c_{21} & \check c_{22} & \check c_{23}
\\ \check c_{31} & \check c_{32} & \check c_{33}
\end{pmatrix} \, ,
\end{equation}
with
\begin{eqnarray}
\check c_{11} & = & 20 N^2 (-1200 + 2252 N^2 - 1096 N^4 + 77 N^6)
\, , \\ \check c_{12} & = & 40 N (-680 + 428 N^2 + 762 N^4 - 195
N^6 + 26 N^8) \, , \nonumber \\ \check c_{13} & = & -20 N (-1360 +
1376 N^2 + 1520 N^4 - 428 N^6 + 13 N^8) \, , \nonumber \\ \check
c_{21} & = & -10 N (-140 - 548 N^2 + 893 N^4 - 190 N^6 + 9 N^8) \,
, \nonumber \\ \check c_{22} & = & 4 (8000 - 17460 N^2 + 11188 N^4
- 2741 N^6 + 114 N^8 + 3 N^{10}) \, , \nonumber \\ \check c_{23} &
= & -32000 + 55240 N^2 - 3232 N^4 - 8306 N^6 + 794 N^8 + 3 N^{10}
\, , \nonumber \\ \check c_{31} & = & 10 N (-2 + 3 N^2) (-220 +
626 N^2 - 126 N^4 + 3 N^6) \, , \nonumber \\ \check c_{32} & = & 4
(-2 + 3 N^2) (-4000 + 7180 N^2 - 2094 N^4 + 252 N^6 + N^8) \, ,
\nonumber \\ \check c_{33} & = & -4 (-2 + 3 N^2) (-4000 + 5730 N^2
- 2629 N^4 + 167 N^6 + N^8) \, . \nonumber
\end{eqnarray}

As an illustration of our differentiation method we re-derive the
second anomalous dimensions in going from the singlet to the
antisymmetric representation: the singlet operators in our
definition have a $g^2$ constant contribution to their two-point
functions arising from the square of the Yukawa additions:
\begin{equation}
\check {\cal S}_{ff} \, = \, \frac{g^2}{4 \pi^2} \, \frac{9N
(N^2-1)}{2(N^2-2)^2(3 N^2-2)^2} \, (8 N - 14 N^3 + 3 N^5 - 4
\xi_f)^2
\end{equation}
On the other hand, the diagonal rescalings of the descendant
operators turn out to be
\begin{equation}
2 \gamma_{1,f} \, a_{2,f} \frac{g^4}{(4 \pi^2)^2} \, = \, \frac{g^4}{(4
\pi^2)^2} \Bigl({\cal Z} (-(\check {\cal B} \otimes \check {\cal
A}^\dagger) \, {\cal G} \, - \, {\cal G} \, (\check {\cal A}
\otimes \check {\cal B}^\dagger)) {\cal Z}^\dagger \Bigr)_{ff} \,
= \, 2 \gamma_{1,f} \, \frac{g^2}{4 \pi^2} \, \check {\cal S}_{ff} \,
.
\end{equation}
(We have taken out the $x$-dependence and a factor of $(4
\pi^2)^{-(J+2)}$.) On forming a difference of the descendant and
$2 \gamma_1 g^2/(4 \pi^2)$ times the singlet two-point functions
these two constants cancel. Matching with the abstract prediction
of harmonic superspace immediately reproduces equation
(\ref{master}).

\section{Conclusions}
\label{concomm}

We have clarified how to compute two-loop anomalous dimensions of
gauge invariant BMN operators using a one-loop calculation
supplemented by differentiation on superspace. The method requires
determining the lowest order mixing of all-scalar and two-fermion
operators, which can be deduced by a set of linear equations. The
second anomalous dimension is found from the square of the two
fermion admixtures. The first problem one encounters in going to
(moderately) higher $J$ is in solving the linear equations. Luckily,
the coefficients in the matrices ${\cal B}, {\cal B}_1, {\cal B}_2$
in (\ref{bmat}), (\ref{beauty}) appear to be either
$O(N^0)$ or $O(N^1)$. It should be possible to find an analytic
solution. Second, higher $J$ is made difficult by the exponential
increase of Wick contractions. Third, to fix the order $g^0$
mixing is a non-linear problem that cannot be solved explicitly in
the general case. The differentiation and the dilation operator
methods share the last feature, of course.

Since the differentiation idea gives one additional loop order for free
it is an obvious avenue of research to try and push the method to the
next loop order at least for low $J$. We hope to obtain information
relevant to the integrability of the spin chain picture.

We note that the values for $\gamma_2$ seem to bear no simple
relation to $\gamma_1$, see also \cite{dilat}. The hope to find
universal formulae for the whole class of operators even at finite
$J$ and $N$ is therefore slim.

Our arguments do certainly rely on a crucial assumption --- namely
that the ${\cal N} = 4$ supersymmetry transformations can more or
less be taken at face value. It is of course a long standing
problem how to justify this: supersymmetry is essentially not
compatible with any known regulator and has to be enforced step by
step using Ward identities.\footnote{We thank M.Bianchi and
G.Rossi for a clarifying discussion on this point.} The
most striking manifestion of the problem relevant to this work is
the occurrence of the Yang-Mills covariant derivative in the
supersymmetry transformation of the spinor fields: supersymmetry
itself is disjoint from gauge symmetry. In a manifest quantum
formulation like ${\cal N} = 1$ one fixes the Wess-Zumino gauge to
eliminate the unphysical fields. Supersymmetry has to be
accompanied by a compensating gauge transformation in order to
conserve the gauge choice \cite{wesbag}, which will eventually lead to the
covariantisation of the derivative. We rather take the point of
view that the covariant derivative is the only possible outcome of
the procedure since otherwise the variation of a gauge singlet
would not be another gauge invariant operator.

The cancellation of the two-fermion part of the $[2,2,0]$ descendant of
${\cal D}_{20}$ is a necessary condition for it to be semishort. We observed
that the operator suffers the general Konishi anomaly and only the
combination with a second term coming from the variation of the covariant
derivative in question will allow the multiplet to be short. The two terms
are in fact of very similar origin since it is parallel transport by the
Yang-Mills covariantised derivative that causes the anomaly.

The supersymmetry variation can act on the connection in a point
splitting regularisation, which is the mechanism originally displayed by
Konishi. It is presumably true that this sort of effect is always
subleading w.r.t. the classical supersymmetry variations, since it
is an essential manner a quantum feature. We will discuss the generalisation
of Konishi's argument to the BMN operators in a forthcoming
publication. When transforming all-scalar operators, the graphs leading
to the one-loop anomaly will cancel in the antisymmetric representation,
while they are present in the BMN singlet. The resulting $g^2$
shift can be traded for a double-fermion admixture. In the
literature it is affirmed that this is an operator identity
\cite{gcr}. In particular, there should not be order $g^2$
reshufflings of the scalars due to the generalised anomaly.
It is perhaps worth investigating whether other anomalous contributions
can arise in the supersymmetry variations of the two-fermion terms.
Careful investigation of the steps of our calculation seems to exclude
this to the given order in the coupling.

The combination of the two contributions in the variation of the
singlets into the two-fermion terms in the antisymmetric
descendants gives a splendid confirmation of our formula for the
generalised Konishi anomaly. We have checked the phenomenon for
$J=0 \ldots 5$ and $SU(N)$ gauge group.

\section {Acknowledgements}
This work is founded in many ways on the collaboration with M. Bianchi,
G. Rossi and Y.S. Stanev that led to our common publication on operator
mixing. During the completion of the article the author benefited
from numerous discussions within the group at Tor Vergata. He would
also like to thank E. Sokatchev for sharing his opinion on the generalised
Konishi anomaly. This work was supported in part by I.N.F.N., by the EC
programs HPRN-CT-2000-00122, HPRN-CT-2000-00131 and HPRN-CT-2000-00148,
by the INTAS contract 99-1-590, by the MURST-COFIN contract 2001-025492
and by the NATO contract PST.CLG.978785.

\section{Group theory}
\label{appnot}

The ${\cal N} = 4$ SYM theory has the fields $\{ \phi_{[AB]}, \,
\psi_{A \, \alpha}, \, \bar \psi^A_{\dot \alpha}, \, A_\mu \}$
transforming under $SU(4)$, i.e. $A,B \in \{1 \ldots
4\}$. The transformation rules are\footnote{In the first line we
mean antisymmetrisation with weight 1.}
\begin{eqnarray}
\delta \phi_{AB} & = & \eta_{[A}^\alpha \psi_{B] \, \alpha} +
\frac{1}{2} \, \epsilon_{ABCD} \, \bar \eta^{[C \, \dot \alpha}
\bar \psi^{D]}_{\dot \alpha} \, , \label{dPhi} \\ \delta \psi_{A
\, \alpha} & = & \eta_A^\beta F_{(\alpha \, \beta)} + g \, \eta_{B
\, \alpha} \, [\phi_{AC}, \phi^{BC}] + \bar \eta^{B \, \dot \beta}
D_{\alpha \, \dot \beta} \, \phi_{AB} \, , \nonumber \\ \delta
\bar \psi^A_{\dot \alpha} & = & \bar \eta^{A \dot \beta} \bar
F_{(\dot \alpha \, \dot \beta)} + g \, \bar \eta^B_{\dot \alpha}
\, [\phi_{BC}, \phi^{AC}] + \eta_B^\beta D_{\dot \alpha \, \beta}
\, \phi^{AB} \, , \nonumber \\ \delta A_\mu & = &  \psi_{I \alpha}
\, (\sigma_\mu)^{\alpha \dot \alpha} \bar \eta^I_{\dot \alpha} +
\eta_{I \alpha} (\sigma_\mu)^{\alpha \dot \alpha} \bar
\psi^I_{\dot \alpha}\, . \nonumber
\end{eqnarray}
The scalar fields obey the reality constraint
\begin{equation}
\overline{(\phi_{AB})} \, = \, \phi^{AB} \, = \, - \frac{1}{2} \,
\epsilon^{ABCD} \, \phi_{CD} \, .
\end{equation}

In order to make contact with the BMN limit it is convenient to
decompose $SU(4)$ into $SO(4)\times U(1)_{J}$. We use $Z =
\phi_{12}$  for the charged singlet and $\phi_2 = \phi_{13}, \,
\phi_3 = \phi_{14}$ to denote the remaining two complex scalars.

The fermions decompose according to
\begin{equation}
\psi^\alpha_A \rightarrow \psi^\alpha_{r \, (+1/2)} \ , \
\psi^\alpha_{\dot{r} \, (-1/2)} \,
\end{equation}
with $A \in \{1 \ldots 4\}, \, r \in \{1,2\}, \, \dot r
\in \{3,4\}$. There is a similar decomposition for the
hermitean conjugate.

So there are the four complex spinors $\psi^\alpha_{r \,(+1/2)}$
and $\bar\psi^{\dot{\alpha}}_{\dot{r} \, (+1/2)}$ with $\Delta
=3/2$ and $J=1/2$, or simply $\Delta -J =1$ and also four complex
spinors $\psi^{\alpha}_{\dot{r} \, (-1/2)}$ and
$\bar\psi^{\dot{\alpha}}_{r \, (-1/2)}$ with $\Delta =3/2$ and
$J=-1/2$, or simply $\Delta-J = 2$.

The supersymmetry charges undergo a similar decomposition and it
turns out to be very convenient to separate them into sets that do
and do not annihilate $Z$. To
this end, define
\begin{eqnarray}
\delta_A \, : & \eta_B \, = \, 0, \, B \neq A \, ; & \bar \eta \,
= \, 0 \, , \\ \bar \delta^A \, : & \bar \eta^B \, = \, 0, \, B \neq A
\, ; & \eta \, = \, 0 \nonumber \, .
\end{eqnarray}
From (\ref{dPhi}) it easy to see that the variations
\begin{equation}
\bar \delta^r \, : \; r \in \{1,2\} \, , \qquad \delta_{\dot
r} \, : \; \dot r \in \{3,4\} \label{goodtrafos}
\end{equation}
preserve $Z$.

Consider first the transformation $\bar \delta^1$:
\begin{equation}
\bar \delta^1 \, Z^J (\phi_2 \bar \phi^2 + \phi_3 \bar \phi^3 + Z
\bar{Z}) = - \bar \eta^{1 \, \dot \alpha}(\phi_2 \bar \psi^3_{\dot
\alpha} + \phi_3 \bar \psi^4_{\dot \alpha} + Z \bar \psi^2_{\dot
\alpha})
\end{equation}
A second application of the same transformation will now act only
on the fermions. Furthermore, the field strength tensor cannot be
generated since $(\bar \eta^1)^2$ has no spin $(0,1)$ part. We
find that $(\bar \delta^1)^2$ acts on the elementary fields in the
singlet operators like (the transformation parameter and a factor
2 have been omitted)
\begin{eqnarray}
& Z \rightarrow 0, & \bar{Z} \rightarrow g \, [\phi_2, \phi_3], \\ &
\phi_2 \rightarrow 0, & \bar \phi^2 \rightarrow - g \, [Z,
\phi_3], \nonumber
\\ & \phi_3 \rightarrow 0, & \bar \phi^3 \rightarrow g \, [Z, \phi_2]
\, . \nonumber
\end{eqnarray}
This is the transformation mainly used in the paper, namely
\begin{equation}
\phi_I \rightarrow 0, \; \bar \phi^I \rightarrow \, \frac{g}{2}
\epsilon^{IJK} [\phi_J, \phi_K]
\end{equation}
where now $I \in \{1,2,3\}$.

More generally, the six components of the antisymmetric $SO(4)$
representation are obtained from the singlets by the scalar part
of
\begin{equation}
\{ \bar \delta^r, \, \bar \delta^s \} \, , \qquad \{ \delta_{\dot
r}, \, \delta_{\dot s} \} \label{comp6}
\end{equation}
and the nine components of the symmetric traceless representation
are found using
\begin{equation}
\{ \bar \delta^r, \, \bar \delta^s \} \, \{ \delta_{\dot r}, \,
\delta_{\dot s} \} \label{comp9}
\end{equation}
(scalar part in both anticommutators).

\section{Technicalities}
\label{technic}

In the antisymmetric and symmetric representations let us choose
$\phi_2$ and $\phi_3$ as impurities. The $N=1$ super Feynman rules
give non-vanishing correlations only with operators of the
conjugate type involving the fields $\bar Z, \bar \phi^2, \bar
\phi^3$. We remark that a symmetric representative with two equal
impurities picks up an extra combinatorical factor of two.

The tree-level mixing between a type I and a type II object is
\begin{eqnarray}
&& \Pi_Z (\phi_2 Z^{J_0} \phi_3 Z^{{J_1}-{J_0}}) \; \; \Pi_{\bar
Z} (\bar \phi^2 \bar Z^{\bar J_1-{\bar J_0}}) (\bar \phi^3 \bar
Z^{\bar J_0}) \label{app11}
\\ & = & \Pi_Z \Pi_{\bar Z} (Z^{J_0} \bar Z^{\bar J_0} Z^{{J_1}-{J_0}} \bar Z^{{\bar J_1}-{\bar J_0}}) +
\frac{c_0^2}{N^2} \Pi_Z \Pi_{\bar Z} (Z^{J_1})(\bar Z^{\bar
J_0})(\bar Z^{{\bar J_1}-{\bar J_0}}) \nonumber
\\ & - & \frac{c_0}{N} \Pi_Z \Pi_{\bar Z} (Z^{J_1} \bar Z^{\bar J_0})(\bar Z^{{\bar J_1}-{\bar J_0}}) - \frac{c_0}{N}
\Pi_Z \Pi_{\bar Z} (Z^{J_1} \bar Z^{{\bar J_1}-{\bar J_0}})(\bar
Z^{\bar J_0}) \nonumber \, .
\end{eqnarray}
It is invariant under ${J_0} \leftrightarrow {J_1}-{J_0}$ and
${\bar J_0} \leftrightarrow {\bar J_1}-{\bar J_0}$.

Let us consider the one-loop Feynman diagrammes. Only matter
exchange graphs contribute. In these a chiral and an antichiral
matter vertex are connected on one leg, leading to the following
three effective vertices:
\begin{equation}
:([\bar \phi^2,\bar \phi^3],[\phi_2, \phi_3]): \qquad :([\bar
Z,\bar \phi^2],[Z, \phi_2]): \qquad :([\bar Z,\bar \phi^3],[Z,
\phi_3]):
\end{equation}
The contraction of the first vertex above on the type II object
gives a commutator $...[\bar Z^{\bar J_0},\bar Z^{{\bar J_1}-{\bar
J_0}}]... = 0$. When dealing with the second vertex it is best to
contract only the $\phi_2,\phi_3$ fields (and c.c.), but to leave
the $\bar Z,Z$ from the vertex untouched. Following \cite{plefkaetal}
the normal ordering can be respected by explicitly subtracting out
a contraction between these fields. After some
simplifications:\footnote{In expressions with $...Z^i \bar Z Z
\bar Z^j...$ we contract, say, the single $\bar Z$ field and
collect the sums into the terms given above. During the process
contractions onto $\Pi_Z$ do occur but they drop out in the end.}
\begin{equation}
\Pi_Z (\phi_2 Z^{J_0} \phi_3 Z^{{J_1}-{J_0}}) \; \; \Pi_{\bar Z}
(\bar \phi^2 \bar Z^{{\bar J_1}-{\bar J_0}}) (\bar \phi^3 \bar
Z^{\bar J_0}) \; \; :([\bar Z,\bar \phi^2],[Z, \phi_2]):
\label{app12}
\end{equation}
\vskip - 1.4 cm
\begin{eqnarray}
= - \Pi_Z \, \Pi_{\bar Z} \, \Bigl( & (Z^{J_0} \bar Z^{\bar
J_0})(Z^{{J_1}-{J_0}} \bar Z^{{\bar J_1}-{\bar J_0}}) + (Z^{J_0}
\bar Z^{{\bar J_1}-{\bar J_0}})(Z^{{J_1}-{J_0}} \bar Z^{\bar J_0})
\nonumber
\\ & - (Z^{J_0})(Z^{{J_1}-{J_0}} \bar Z^{\bar J_1}) - (Z^{{J_1}-{J_0}})(Z^{J_0} \bar Z^{\bar J_1}) \Bigr)
\nonumber
\end{eqnarray}
The result is symmetric under ${J_0} \leftrightarrow {J_1}-{J_0}$
and ${\bar J_0} \leftrightarrow {\bar J_1}-{\bar J_0}$ separately.
The remaining effective vertex yields the expression above with
both index exchanges, an equal contribution.

Next, the one-loop mixing between two type II operators shares the
feature that the first effective vertex does not contribute. The
second vertex yields
\begin{equation}
\Pi_Z (\phi_2 Z^{J_0})(\phi_3 Z^{{J_1}-{J_0}}) \; \; \Pi_{\bar Z}
(\bar \phi^2 \bar Z^{{\bar J_1}-{\bar J_0}}) (\bar \phi^3 \bar
Z^{\bar J_0}) \; \; :([\bar Z,\bar \phi^2],[Z, \phi_2]):
\label{app13}
\end{equation}
\vskip -1.4 cm
\begin{eqnarray}
= 2 \Pi_Z \Pi_{\bar Z} \Bigl( (Z^{J_1} \bar Z^{\bar J_1}) -
(Z^{J_0} \bar Z^{{\bar J_1}-{\bar J_0}} Z^{{J_1}-{J_0}} \bar
Z^{\bar J_0}) \Bigr) \, . \nonumber
\end{eqnarray}
This expression is again symmetric under both ${J_0}
\leftrightarrow {J_1}-{J_0}$ and ${\bar J_0} \leftrightarrow {\bar
J_1}-{\bar J_0}$. The third vertex gives an equal contribution.

Let us now turn to the one-loop mixing in the singlet sector. We
start by discussing the mixing of type I with type II operators.
As above, in non-vanishing one-loop graphs the interaction must
involve the impurities. We may divide the calculation into two
sectors: first, one impurity of each of the operators in the
two-point function is involved. The total contribution of such
graphs is 16 times the r.h.s. of equation (\ref{app12}). Second,
the interaction is only between the impurities. The contribution
from this sector is equal but of opposite sign, hence there is
exact cancellation.

This pattern is repeated in the mixing of type II with type II
operators: we find 16 times the r.h.s. of equation (\ref{app13})
from the first sector and its negative from the second. It is easy
to check that type II operators do not mix with the type III
either. We arrive at the conclusion that type II singlet operators
are one-loop protected.

Next we address the non-vanishing two-point functions. The
one-loop mixing of a type III with another type III is via a
Yang-Mills exchange. We find:
\begin{equation}
\langle \Pi_Z (Z^{J_1} \bar Z) \; \Pi_{\bar Z} (\bar Z^{\bar J_1}
Z) \rangle_{g^2} \, = \, 2 \Pi_Z \Pi_{\bar Z} \Bigl(
(Z^{J_1})(\bar Z^{\bar J_1}) - ()(Z^{J_1} \bar Z^{\bar J_1})
\Bigr)\label{app14}
\end{equation}
For the mixing of a type I operator with a type III we find only
one sort of matter exchange diagramme, here calculated for the
first term of the type I singlet:
\begin{equation}
\Pi_Z (\phi_2 Z^{J_0} \bar \phi^2 Z^{{J_1}-{J_0}}) \; \; \Pi_{\bar
Z} (\bar Z^{\bar J_1} Z) \; \; :([\phi_2, Z] [\bar \phi^2 \bar
Z]): \label{app15}
\end{equation}
\vskip - 1.4 cm
\begin{eqnarray}
 = \, \Pi_Z \Pi_{\bar Z} \bigl( & (Z^{{J_0}+1}) (Z^{{J_1}-{J_0}}
\bar Z^{\bar J_1}) + (Z^{{J_1}-{J_0}+1})(Z^{J_0} \bar Z^{\bar
J_1}) \nonumber \\ & - (Z^{J_0})(Z^{{J_1}-{J_0}+1} \bar Z^{\bar
J_1}) - (Z^{{J_1}-{J_0}})(Z^{{J_0}+1} \bar Z^{\bar J_1}) \bigr)
\nonumber
\end{eqnarray}
This is symmetric under ${J_0} \leftrightarrow {J_1}-{J_0}$ so
that the other three terms of the type I singlet all add an equal
contribution.

Third, the type I / type I mixing matrix is
\begin{equation}
\langle \Pi_Z ( (\phi_2 Z^{J_0} \bar \phi^2 Z^{{J_1}-{J_0}}) +
\ldots ) \; \Pi_{\bar Z} ( (\phi_2 \bar Z^{\bar J_0} \bar \phi^2
\bar Z^{{\bar J_1}-{\bar J_0}}) + \ldots ) ) \rangle_{g^2}
\label{app16}
\end{equation}
\vskip -1.4 cm
\begin{eqnarray}
= 8 \, \Pi_Z \Pi_{\bar Z} \bigl( & \; \; \, (Z^{{J_0}+1} \bar
Z^{\bar J_0})(Z^{{J_1}-{J_0}} \bar Z^{{\bar J_1}-{\bar J_0}+1}) +
(Z^{J_0} \bar Z^{{\bar J_0}+1}) (Z^{{J_1}-{J_0}+1} \bar Z^{{\bar
J_1}-{\bar J_0}}) \nonumber
\\ & - (Z^{J_0} \bar Z^{\bar J_0})(Z^{{J_1}-{J_0}+1} \bar Z^{{\bar J_1}-{\bar J_0}+1}) - (Z^{{J_0}+1} \bar
Z^{{\bar J_0}+1})(Z^{{J_1}-{J_0}} \bar Z^{{\bar J_1}-{\bar J_0}})
\nonumber \\ & + ({\bar J_0} \leftrightarrow {\bar J_1}-{\bar
J_0}) \bigr) \nonumber \, .
\end{eqnarray}

\end{document}